\documentclass[twocolumn,preprintnumbers,nofootinbib]{revtex4-1}
\usepackage{graphicx}
\usepackage{dcolumn}
\usepackage{bm}
\usepackage{amsmath,amssymb,amsfonts}
\usepackage{latexsym}
\usepackage{bbm}
\usepackage{url} 
\usepackage{color}
\usepackage{soul}
\usepackage{multirow}
\usepackage[caption=false]{subfig}
\usepackage{braket}

\begin{document}
\captionsetup[subfigure]{labelformat=empty}
\preprint{ACFI-T22-04, INT-PUB-22-009, LA-UR-21-32420}

\title{Next-to-leading order scalar contributions to $\mu\to e$ conversion }

\author{Vincenzo Cirigliano$^{1,2}$}
\email{cirigv@uw.edu}
\author{Kaori Fuyuto$^1$}
\email{kfuyuto@lanl.gov}
\author{Michael J. Ramsey-Musolf$^{~3,4,5}$}
\email{mjrm@physics.umass.edu}
\author{Evan Rule$^6$}
\email{erule@berkeley.edu}
\affiliation{$^1$ Theoretical Division, Los Alamos National Laboratory, Los Alamos, NM 87545, USA}
\affiliation{$^2$ Institute for Nuclear Theory, University of Washington, Seattle WA 98195, USA}
\affiliation{$^3$ Tsung-Dao Lee Institute and  School of Physics and Astronomy, Shanghai Jiao Tong University, 800 Dongchuan Road, Shanghai 200240}
\affiliation{$^4$ Amherst Center for Fundamental Interactions, Department of Physics,
University of Massachusetts Amherst, MA 01003, USA}
\affiliation{$^5$ Kellogg Radiation Laboratory, California Institute of Technology, Pasadena, CA 91125 USA}
\affiliation{$^6$ Department of Physics, University of California, Berkeley, CA 94720, USA}

\bigskip

\date{\today}

\begin{abstract}

Within a class of models in which lepton flavor violation is induced dominantly by scalar particle exchanges, 
we estimate  the $\mu \to e$ conversion rate in several nuclei. We include  next-to-leading order (NLO) terms in the one- and two-nucleon interactions in chiral effective theory,  rectifying some incorrect results in the previous literature. We provide an uncertainty budget for the conversion rates and we find that NLO contributions affect the amplitudes at the level of $10\%$, which could be larger than the  uncertainty  on the leading order couplings, dominated by the strange and 
non-strange nucleon sigma terms. 
We  study the  implications of our results for testing Higgs-mediated CLFV in the future by combining results from various experimental searches, such as $\mu \to e$ conversion in multiple target nuclei and $\mu \to e \gamma$.
\end{abstract}

\maketitle

\section{Introduction}
\label{sec:intro}

Lepton flavor violating processes involving charged leptons (CLFV) are among the theoretically cleanest probes of physics beyond the Standard Model (BSM).  The minimal model extending the Standard Model with neutrino masses predicts  CLFV amplitudes proportional to  $\Delta m_\nu^2/m_W^2$ \cite{Petcov:1976ff,Marciano:1977wx,Lee:1977qz,Lee:1977tib}, 
leading to branching ratios forty orders of magnitude below current experimental sensitivity and hence a huge discovery window.   Moreover, CLFV processes also test the origin of flavor breaking in the lepton sector, ultimately related to the structure of neutrino mass matrices. 
So far, the strongest experimental bounds on CLFV have been given on $\mu\to e$ transitions. The current limit on the branching ratio of $\mu\to e\gamma$ is $B_{\mu\to e\gamma}<4.2\times 10^{-13}$ at $90\%$ C.L. \cite{MEG:2016leq}, while that of $\mu\to e$ conversion in gold is $B_{\mu\to e}({\rm Au})<7\times 10^{-13}$ at $90\%$ C.L. \cite{SINDRUMII:2006dvw}. The next-generation searches aim to achieve higher sensitivities. For example, the MEG II experiment at the Paul Scherrer Institute (PSI) is expected to reach $B_{\mu\to e\gamma}< 6\times 10^{-14}$ \cite{MEGII:2018kmf, MEGII:2021fah}. The Mu2e  experiment at Fermilab and the COherent Muon to Electron Transition (COMET) experiment at Japan Proton Research Complex (J-PARC) plan to increase their sensitivity reaches by four orders of magnitude, i.e. to the level of $B_{\mu\to e}< O(10^{-17})$ \cite{Mu2e:2014fns, Mu2e:2018osu, COMET:2018auw}.

Compared to the non-hadronic decay $\mu \to e \gamma$, the process for $\mu\to e$ conversion  occurs in nuclei, wherein a muon is trapped to form a muonic atom. 
The $\mu \to e$ conversion process has the potential to discern signs of various BSM physics (for early studies see Ref.~\cite{Marciano:1977cj}), since it can be generated by not only photonic dipole operators but also non-photonic contact four-fermion interactions involving two leptons and two quarks, with various Lorentz structures. 
Thus, estimations of the conversion process involve multiple 
scales (hadronic, nuclear, atomic)  and require careful treatments. 
In light of the expected improvements in experimental sensitivity, 
in order to maximize the constraining power (in case of null signal) or 
the model-diagnosing power (in case of discovery), 
accurate theoretical predictions within various classes of models are  desirable. 

The calculation of $\mu \to e$ conversion rates has a long history, 
starting with the pioneering work of Weinberg and Feinberg~\cite{Weinberg:1959zz}, in which the electron wavefunction was 
taken as a plane wave and muon wavefunction was approximated to a constant. 
It was later realized that relativistic effects can be important in medium and heavy nuclei~\cite{Shanker:1979ap}, and detailed calculation of the conversion rate including relativistic lepton wavefunctions was performed in Refs.~\cite{Czarnecki:1998iz,Kitano:2002mt}.
The conversion rates in various nuclei and for dipole, scalar, and vector operators have been calculated in \cite{Kitano:2002mt}, 
where the uncertainty induced by neutron and proton densities 
in the ground state is also estimated. 
The discriminating power of various interactions in $\mu\to e \gamma$ and $\mu\to e$ conversion was further assessed in \cite{Cirigliano:2009bz}, where the uncertainty from scalar form factors was also discussed. 
The effect of nucleon spin-dependent operators was first discussed in Refs.~\cite{Cirigliano:2017azj,Davidson:2017nrp} and 
effective field theory studies including renormalization group evolution, 
along with their phenomenological implications, have appeared~\cite{Crivellin:2017rmk,Davidson:2018kud,Davidson:2020hkf}.
Very recently, a nuclear-level effective theory for 
$\mu \to e$ conversion was developed~\cite{Rule:2021oxe}. 

Among the spin-independent operators, the largest theoretical uncertainties 
arise in the scalar sector. 
An improved treatment of scalar  matrix elements based on SU(2) chiral perturbation theory (ChPT)  was introduced in Ref.~\cite{Crivellin:2014cta}.  
More  recently, in the framework of SU(2) ChPT, the impact of next-to-leading order (NLO) nucleon interactions 
induced by quark-level scalar densities  was discussed in  Ref.~\cite{Bartolotta:2017mff}. The NLO contributions involve 
both single-nucleon scalar form factors and 
two-nucleon interactions, which were reduced  in Ref.~\cite{Bartolotta:2017mff} 
to an effective one-nucleon interaction by performing an average of the interaction over a Fermi gas model. It was found that, when scalar operators are the dominant sources for the conversion process, the NLO interactions could bring destructive contributions relative to LO contributions and significantly reduce the branching ratio.

Inspired by those previous studies, we revisit the calculation of branching ratios of $\mu \to e$ conversion in several nuclei including the NLO nucleon interactions induced by scalar quark densities. 
Our analysis is performed in a model-independent way, i.e. the Standard Model Effective Field Theory (SM-EFT)~\cite{Weinberg:1979sa,Wilczek:1979hc,Buchmuller:1985jz,Grzadkowski:2010es}, focusing on a particular class of SM-EFT operators: photonic dipole and scalar four-fermion operators. This setup is well motivated by BSM models with heavy scalar particles such as Two-Higgs Doublet Model and Leptoquark Models. As a byproduct of our analysis, we correct two errors appearing in \cite{Bartolotta:2017mff}: (1) we eliminate two un-physical contributions to the overlap integrals (denoted by $\tau^{(2,3)}$);  
(2) we provide a corrected expression and corresponding 
numerical result for the effective one-nucleon interaction resulting from the average over the Fermi gas model. 
We also develop a new method to include the momentum-dependence of the scalar form factor in the overlap integrals. 
With these results at hand,  we  assess the NLO contributions to the conversion process and discuss the current hadronic and nuclear uncertainties. 
We discuss the prediction of this scalar-dominance model for the ratio of 
$\mu \to e \gamma$ over $\mu \to e$ conversion in $^{27}{\rm Al}$ and 
for the ratio of conversion rates in different target nuclei. 
Finally, we apply the analysis to 
a simple model in which CLFV is mediated by CLFV Yukawa couplings of the SM Higgs,  which realizes the current setup. 

The paper is organized as follows. In Section~\ref{sec:operator} we set up the EFT framework, 
starting from quark-level interactions and matching to nucleon-level interactions. 
In Section~\ref{sec:transition} we present the results for the overlap integrals and the $\mu \to e$ conversion rate, 
including NLO chiral effects. 
In Section~\ref{sec:results} we discuss the implication for scalar-mediated CLFV first in the EFT setup and subsequently 
in a model with `minimal' Higgs-mediated CLFV, i.e. an extension of the Standard Model in which the only new interactions are 
CLFV Yukawa couplings of the Higgs to leptons. 
We present our conclusions in Section~\ref{sec:conclusion} and relegate some technical details to the Appendices.

\section{Effective interactions: from quarks to nucleons}
\label{sec:operator}
We assume in this work that the BSM physics responsible for CLFV originates at energies above the electroweak scale. 
In this case contributions from any BSM physics are captured by effective operators 
expressed in terms of SM fields, with appropriate couplings that contain 
information about the underlying model -- this  is the 
SM-EFT framework.
We restrict our attention to a particular class of SM-EFT operators mediating CLFV transitions, namely the  
ones mediated by heavy scalar particles, including the SM Higgs itself. 
As our goal is to assess  the uncertainties in scalar-mediated CLFV, we take as starting point  
below the electroweak scale the following effective Lagrangian (for a complete set of operators see Refs.~\cite{Kitano:2002mt, Cirigliano:2009bz}) 
\begin{align}
{\cal L}_{\rm eff}=&
-\frac{1}{\Lambda^2}
\sum_{\alpha=L,R} 
\bigg[  C_{D\alpha}\ m_{\mu} \ \bar{e}\sigma^{\lambda \nu}P_\alpha\mu \,  F_ {\lambda \nu}
\nonumber\\
&+\sum_{q=u,d,s,c,b,t}  C^{(q)}_{S\alpha} \  G_F m_\mu  m_q  \  \bar{q} q  \, \bar{e}P_\alpha \mu 
\nonumber\\
&  + \ \  C_{G \alpha} \  G_F m_\mu  \alpha_s \  G^a_{\lambda \nu}  G^{a \, \lambda \nu}  \bar{e}P_\alpha \mu 
+{\rm h.c }\bigg], 
\label{effective_operator}
\end{align}
where  $F_{\mu\nu}$ is the field strength of the photon, $P_{L,R} = (1 \mp \gamma_5)/2$ are the chirality projectors,  $\Lambda$ represents the new physics scale,  and the Wilson coefficients $C_{D\alpha}$, $C_{S\alpha}^{(q)}$  are dimensionless.

With this normalization, the chirality flip in lepton and quark bilinears is accompanied by a muon or a quark mass insertion, respectively.\footnote{ 
The scalar interactions have an additional factor $G_Fm_{\mu}m_q$ compared to the definition in \cite{Bartolotta:2017mff}.} 
This choice comes without loss of generality  and simplifies many intermediate steps in the analysis. 
Finally, the factors of $m_q$ and $\alpha_s$ multiplying the quark scalar bilinears and the gluonic operator ensure that the corresponding Wilson coefficients do not run under QCD renormalization.  After integrating out the heavy quarks, at the GeV scale the effective Lagrangian takes the form of Eq.~(\ref{effective_operator}), 
with $q= u,d,s$ and~\cite{Shifman:1978zn}
\begin{align}
C_{G\alpha} \to C_{G\alpha} -1/(12 \pi) \sum_{Q=c,b,t} C_{S \alpha}^{(Q)}\ \ \ .
\label{heavy_qaurks}
\end{align}

The scalar  quark operators in Eq. (\ref{effective_operator}) induce single- and multi-nucleon momentum-dependent operators 
at low-energy, which  eventually lead to nuclear transitions.  
The form of one- and two-nucleon operators has been derived  to NLO in both SU(3) ChPT~\cite{Cirigliano:2012pq} 
and SU(2) ChPT~\cite{Crivellin:2013ipa,Crivellin:2014cta,Korber:2017ery,Bartolotta:2017mff}, and we will work here in the SU(2) case.  
Making   the following identifications (recall $\alpha \in \{L,R\}$ is a chirality label for the lepton bilinear appearing in the scalar operators)
\begin{widetext}
\begin{subequations}
\begin{align}
\langle N (\mathbf{k}^\prime)  |    \, C_{S\alpha}^{(u)} m_u \bar u u + C_{S\alpha}^{(d)} m_d \bar d d  \, |  N (\mathbf{k} )\rangle 
& \to \ 
\bar{N}^\prime  \, J^{(1)}_{ud,\alpha}  (\mathbf{q})     \, N
\ \ 
\\  
\langle N( \mathbf{k}_1^\prime) N (\mathbf{k}_2^\prime)   |    \, C_{S\alpha}^{(u)} m_u \bar u u + C_{S\alpha}^{(d)} m_d \bar d d  \, | 
N( \mathbf{k}_1) N (\mathbf{k}_2)   \rangle 
& \to \  
\bar{N}_1^\prime \bar{N}_2^\prime  \,   
J^{(2)} _{ud,\alpha}  (\mathbf{q}_1, \mathbf{q}_2)   
 \, N_1 N_2
\\
\langle N (\mathbf{k}^\prime)  |    \, C_{S\alpha}^{(s)} m_s \bar s s  \, |  N (\mathbf{k} ) \rangle 
& \to \ 
\bar{N}^\prime \, J^{(1)}_{s,\alpha}  (\mathbf{q})    \, N
\\
\langle N (\mathbf{k}^\prime)  |    \, C_{G \alpha} \alpha_s   G^a_{\lambda \nu}  G^{a \, \lambda \nu}   \, |  N (\mathbf{k} ) \rangle 
& \to \ 
\bar{N}^\prime \,       J ^{(1)}_{G,\alpha}  (\mathbf{q})   \, N
\\
\langle N( \mathbf{k}_1^\prime) N (\mathbf{k}_2^\prime)   |    \,    C_{G \alpha} \alpha_s   G^a_{\lambda \nu}  G^{a \, \lambda \nu}   \, | 
N( \mathbf{k}_1) N (\mathbf{k}_2)   \rangle 
& \to \  
\bar{N}_1^\prime \bar{N}_2^\prime  \,     
  J^{(2)} _{G,\alpha}  (\mathbf{q}_1, \mathbf{q}_2)       
   \, N_1 N_2
\end{align}
\end{subequations} 
\end{widetext}
where $\mathbf{q}= \mathbf{k}^\prime - \mathbf{k}$, $\mathbf{q}_i = \mathbf{k}_i^\prime - \mathbf{k}_i$ 
and $N$ denotes  the non-relativistic spinors for the nucleon doublet, one finds that the hadronic currents are expressed by the Wilson coefficients in Eqs. (\ref{effective_operator}), (\ref{heavy_qaurks}) and (\ref{isoscalar_vector}) and hadronic parameters in Eq. (\ref{hadronic_parameters}) ~\cite{Crivellin:2013ipa,Crivellin:2014cta,Korber:2017ery}
\begin{widetext}
\begin{subequations}
\label{eq:currents}
\begin{eqnarray}
J^{(1)} _{ud,\alpha}  (\mathbf{q})    &=&
 \left[ \sigma_{\pi N}  -\frac{3 m_{\pi}^3 {g}^2_A}{64\pi {f}^2_{\pi}}  \, F(\mathbf{q}^2/m_\pi^2) \right] 
 C_{S\alpha}^{(0)}
- \frac{\delta m_N}{4} \tau_3 
\, C_{S\alpha}^{(1)}  
\\
 J^{(2)} _{ud,\alpha}  (\mathbf{q}_1, \mathbf{q}_2)    &=&
- \frac{ g_A^2 m_\pi^2}{4 f_\pi^2} 
 \frac{ \mathbf{\sigma}_1 \cdot \mathbf{q}_1 \, \mathbf{\sigma}_2 \cdot \mathbf{q}_2}{ (\mathbf{q}_1^2 + m_\pi^2) (\mathbf{q}_2^2 + m_\pi^2)}
 \tau_1 \cdot \tau_2 
 \, C_{S\alpha}^{(0)}
 \\
 J^{(1)} _{s,\alpha}  (\mathbf{q})    &=&
  \left( \sigma_s - \dot{\sigma}_s  \mathbf{q}^2 \right)
  \, C_{S\alpha}^{(s)} 
 \\
  J^{(1)} _{G,\alpha}  (\mathbf{q})   &=& - \frac{8 \pi}{9} C_{G \alpha} \, 
  \left( m_N 
  - \left[ \sigma_{\pi N}  -\frac{3 m_{\pi}^3  {g}^2_A}{64\pi {f}^2_{\pi}}  \, F(\mathbf{q}^2/m_\pi^2) \right] 
  + \frac{\delta m_N}{2} \tau_3 
  -  \left( \sigma_s - \dot{\sigma}_s  \mathbf{q}^2 \right)  
  \right)
 \\
  J^{(2)} _{G,\alpha}  (\mathbf{q}_1, \mathbf{q}_2)    &=& 
  - \frac{8 \pi}{9} C_{G \alpha} \ 
 \frac{ g_A^2 m_\pi^2}{4 f_\pi^2} 
 \frac{ \mathbf{\sigma}_1 \cdot \mathbf{q}_1 \, \mathbf{\sigma}_2 \cdot \mathbf{q}_2}{ (\mathbf{q}_1^2 + m_\pi^2) (\mathbf{q}_2^2 + m_\pi^2)}
 \tau_1 \cdot \tau_2  ~. 
\end{eqnarray} 
\end{subequations}
\end{widetext}
The isoscalar and isovector combinations of scalar  Wilson coefficients are given by:
\begin{subequations}
\label{eq:WilsonC}
\begin{eqnarray}
C_{S\alpha}^{(0)}    &=& \frac{C_{S\alpha}^{(u)} ( 1 - \epsilon) + C_{S\alpha}^{(d)} ( 1 + \epsilon)  }{2} 
\\
C_{S\alpha} ^{(1)}    &=&   C_{S\alpha}^{(u)} \left( 1 - \frac{1}{\epsilon} \right) 
+ C_{S\alpha}^{(d)} \left( 1 + \frac{1}{\epsilon} \right) 
 ~. 
\end{eqnarray}
\label{isoscalar_vector}
\end{subequations}
The single-nucleon scalar form factor is given by 
\begin{eqnarray}
F(x) &=&  \frac{2 + x}{\sqrt{x}}  \, {\rm ArcCot} \left( \frac{2}{\sqrt{x}} \right) - 1
\\
&\simeq &\frac{5}{12}x -\frac{7}{240} x^2+\cdots.~. 
\end{eqnarray}
We note that  the first order Taylor expansion provides a representation of the full expression 
accurate to 2\% for $x \leq 1$ and to $5\%$ for $ 1 < x \leq 2$. 
This means that it is quite safe to use the linear term in our nuclear analysis. 
In Appendix. \ref{app:transition}, we show how the momentum-transfer expansion corresponds to  derivative operators acting on nucleon density functions. 

The hadronic scalar current defined in this way carries uncertainties due to both input parameters and higher order terms in the chiral expansion. 
Higher order terms in the momentum-independent part of the current are effectively resummed by using the 
physical values of $\sigma_{\pi N}$, $\sigma_s$, and $\delta m_N$.  
The hadronic inputs entering Eqs.~(\ref{eq:currents}) are defined as
\begin{subequations}
\begin{eqnarray}
\sigma_{\pi N} &=&  \frac{1}{2} \langle N | \,  (m_u + m_d) (\bar u u  + \bar d d)  \, | N \rangle
\\
\sigma_s &=& \langle N | \, m_s \bar s s \, | N \rangle 
\\
\epsilon &=& \frac{m_d- m_u}{m_d+ m_u}  
\\
\delta m_N &=& (m_n - m_p)_{\rm strong}~.
\end{eqnarray}
\label{hadronic_parameters}
\end{subequations} 
For the sigma term, we use as  baseline input the analysis of the Roy-Steiner equations of Ref.~\cite{Hoferichter:2015dsa}, namely 
$\sigma_{\pi N} =  59.1(3.5)$~MeV.  
This value is in tension with the  (currently) more uncertain  lattice QCD calculations (see \cite{Aoki:2021kgd}  and references therein) 
which indicate $\sigma_{\pi N} = 65(13)$~MeV (with dynamical charm quark~\cite{Alexandrou:2014sha}) 
and $\sigma_{\pi N} = 40(4)$~MeV (no dynamical charm quark~\cite{Durr:2011mp,Durr:2015dna,Yang:2015uis}).
However, the recent lattice QCD analysis of Ref.~\cite{Gupta:2021ahb} suggests that inclusion of excited state effects reconciles the tension, so we 
shall use the input from Ref.~\cite{Hoferichter:2015dsa}. 
For the strange sigma term we use the lattice QCD average~\cite{Aoki:2021kgd}
$\sigma_s = 41(9)$MeV (with dynamical charm~\cite{Freeman:2012ry}), 
while for the slope we will take $\dot \sigma_s = 0.3(2)~{\rm GeV}^{-1}$~\cite{Hoferichter:2012wf}. 
For the strong-isospin contribution to the nucleon mass splitting we take the lattice QCD determination $\delta m_N = 2.32 (17)$~MeV from Ref.~\cite{Brantley:2016our}, 
consistent with the earlier lattice calculation of Ref.~\cite{Borsanyi:2014jba}.  Finally, we take $\epsilon = 0.365(23)$ from the FLAG average~\cite{Aoki:2021kgd}.

Higher chiral orders in the momentum dependence of the single-nucleon form factor  are expected to be  sizable. 
In fact, a  comparison of the NLO heavy baryon ChPT prediction~\cite{Bernard:1992qa}  with a recent dispersive determination~\cite{Hoferichter:2012wf}
indicates  that the NLO result accounts for about 60\% of the dispersive result. 

Finally, note that the one- and two-nucleon amplitudes
$\mu N(\mathbf{k}) \to  e N(\mathbf{k}^\prime) $ 
and 
$\mu N(\mathbf{k_1}) N(\mathbf{k_2})  \to  e N(\mathbf{k}_1^\prime) N(\mathbf{k}_2^\prime) $ 
take the form  
\begin{subequations}
\begin{eqnarray}
{\cal A}^{(1)}  &=& - \frac{G_F m_\mu}{\Lambda^2}  \,  \sum_{\alpha = L,R} \bar{N}^\prime   J^{(1)}_\alpha N \   \langle  \bar{e} P_\alpha \mu \rangle 
\\
{\cal A}^{(2)}  &=& - \frac{G_F m_\mu}{\Lambda^2}  \,  \sum_{\alpha = L,R} \bar{N}_1^\prime \bar{N}_2^\prime  J^{(2)}_\alpha N_1 N_2
 \   \langle  \bar{e} P_\alpha \mu \rangle 
 \qquad 
\end{eqnarray}
\end{subequations} 
where $\langle \bar e P_\alpha \mu \rangle$ denotes the leptonic amplitude and the  physically 
relevant combinations of hadronic scalar currents are 
\begin{subequations}
\begin{eqnarray}
J^{(1)}_\alpha &=& J^{(1)}_{ud,\alpha} + J^{(1)}_{s,\alpha} +  J^{(1)}_{G,\alpha}  
\\
J^{(2)}_\alpha &=& J^{(2)}_{ud,\alpha} +  J^{(2)}_{G,\alpha}  ~, 
\end{eqnarray}
\end{subequations} 
which take the form 
\begin{widetext}
\begin{subequations}
\label{eq:physical_currents}
\begin{eqnarray}
J^{(1)}_\alpha  (\mathbf{q})   &=&   
\left[ \sigma_{\pi N}  -\frac{3 m_{\pi}^3 {g}^2_A}{64\pi {f}^2_{\pi}}  \, F(\mathbf{q}^2/m_\pi^2) \right] 
\left( C_{S\alpha}^{(0)} + \frac{8 \pi}{9} C_{G\alpha} \right)
- \frac{\delta m_N}{4} \tau_3 
\left( C_{S\alpha}^{(1)} + \frac{16 \pi}{9} C_{G\alpha} \right)
\label{eq:physical_currents_1}
\nonumber 
\\
&+&   \left( \sigma_s - \dot{\sigma}_s  \mathbf{q}^2 \right)
  \, \left( C_{S\alpha}^{(s)}  + \frac{8 \pi}{9} C_{G\alpha} \right)
- \frac{8 \pi}{9} C_{G\alpha} m_N 
\\
J^{(2)}_\alpha    (\mathbf{q}_1, \mathbf{q}_2)   &=& 
-  \frac{ g_A^2 m_\pi^2}{4 f_\pi^2} 
 \frac{ \mathbf{\sigma}_1 \cdot \mathbf{q}_1 \, \mathbf{\sigma}_2 \cdot \mathbf{q}_2}{ (\mathbf{q}_1^2 + m_\pi^2) (\mathbf{q}_2^2 + m_\pi^2)}
 \tau_1 \cdot \tau_2 
 \, \left(  C_{S\alpha}^{(0)} + \frac{8 \pi}{9} C_{G\alpha} \right)~.
\end{eqnarray}
\end{subequations} 
\end{widetext}
If the gluonic operator is  sourced only by integrating out the heavy quarks, 
one has the relation 
\begin{equation}
\frac{8 \pi}{9} C_{G\alpha} =  - \frac{2}{27} \sum_{Q=c,b,t}  C_{S\alpha}^{(Q)}~. 
\end{equation}

{\renewcommand{\arraystretch}{1.5}
\begin{table}
\begin{tabular}{lcccc}
\hline
 & $^{27}_{13}$Al & $^{48}_{22}$ Ti & $^{197}_{79}$Au & $^{208}_{82}$Pb\\
\hline
\hline
$q_T$ (MeV) & 104.97 & 104.27 & 95.61 & 95.10 \\
$k_F$ (MeV) & 238 & 255 & 265 & 265\\
$R_p$ (fm) & 3.05 & 3.843 & $6.55$ & 6.624\\
$R_n$ (fm) & $3.18\pm 0.19$ & 3.843 & $6.83\pm 0.1$ & $6.93\pm 0.09$\\
$a$ (fm) & 0.535 & 0.588 & 0.522 & 0.549\\
\hline
$f_\mathrm{eff,FGA}^{SI}$ & $0.43^{+0.03}_{-0.22}$ & $0.49^{+0.03}_{-0.25}$ & $0.55^{+0.03}_{-0.28}$ & $0.55^{+0.03}_{-0.28}$\\
$f_\mathrm{eff,NSM}^{SI}$ & 0.18 & 0.18 & - & -\\
\hline
\end{tabular}
\caption{Input parameters and resulting values of the effective one-body operator coupling parameter $f^{SI}_\mathrm{eff}$ for the four nuclei of interest: $q_T$ is the magnitude of the three-momentum transfer, computed via Eq. (\ref{eq:q_val}), $k_F$ is the nuclear Fermi momentum obtained by linear interpolation in $A$ between the values measured in \cite{Moniz:1971mt}, $R_p$ ($R_n$) and $a$ are the parameters of the proton (neutron) density profile, $f_\mathrm{eff,FGA}^{SI}$ is the value of the spin-independent form factor obtained via the Fermi gas average, and $f^{SI}_\mathrm{eff,NSM}$ is the value implied by the nuclear shell model calculation (without additional correlation function). The upper uncertainty of $f^{SI}_\mathrm{eff,FGA}$ is due to the error incurred in the momentum average over $\bar{k}$ and the uncertainty in the Fermi momentum ($\pm 5$ MeV) whereas the lower uncertainty reflects the expectation that the FGA result tends to overestimate the strength of the operator by roughly a factor of 2. See discussion in Appendix 
 \ref{app:two_nucleon} and  \ref{app:1body_eff}  for more details. }
\label{tab:effective_op}
\end{table}
}

The NLO  two-nucleon contribution  can be  reduced to an effective single-nucleon operator 
by averaging the two-nucleon operator over a Fermi gas model of the target nucleus. In this approximation, the effect of 
$J^{(2)}_\alpha$ is captured by the shift 
\begin{equation}
\sigma_{\pi N} \to \sigma_{\pi N}  -\frac{3  g_A^2  m_{\pi}^2 k_F }{64\pi {f}^2_{\pi}}    \, f^{SI}_\mathrm{eff}
\end{equation}
in the first term in   $J^{(1)}_\alpha$  in Eq.~(\ref{eq:physical_currents_1}), where $k_F$ is the Fermi momentum in the target nucleus and 
$f^{SI}_\mathrm{eff}$ is the effective single-nucleon coupling resulting from averaging over the Fermi gas model. 
Although this procedure was carried out in \cite{Bartolotta:2017mff}, an error in that calculation resulted in incorrect results for the effective single-nucleon form factors $f^{SI}(\bar{q},\bar{k})$ and $f^{SD}(\bar{q},\bar{k})$ which propagated to all values obtained from these form factors. In Appendix \ref{app:1body_eff}, we present the corrected expressions. The corrections significantly affect the values obtained after momentum-averaging, reducing the overall magnitudes by roughly a factor of two and leading to
$f^{SI}_\mathrm{eff}=0.43^{+0.03}_{-0.22}$ and $f^{SD}_\mathrm{eff}=0.43^{+0.03}_{-0.22}$.

As discussed in Ref.\cite{Bartolotta:2017mff}, the effective coupling $f^{SI}_\mathrm{eff}$ obtained through the Fermi gas average is likely an overestimate of the underlying two-nucleon contribution. This expectation is based on a study of nuclear anapole moments \cite{Haxton:2001zq} where, in addition to the Fermi gas average, nuclear shell model wave functions were used to directly evaluate a variety of two-nucleon currents -- none of them the operator of present concern -- in the nuclei $^{133}$Cs and $^{205}$Tl. Across all the currents tested, the Fermi gas average tended to overestimate the two-nucleon contribution by 2-3 times compared to the shell model.

To verify that this behavior persists in the present case, we evaluated the two-nucleon operator using shell model wave functions for two of our nuclei of interest, $^{27}$Al and $^{48}$Ti. Details of this calculation are presented in Appendix \ref{app:two_nucleon}. Fully correlated shell model wave functions were generated for $^{27}$Al and $^{48}$Ti using the configuration-interaction code BIGSTICK \cite{Johnson:2013bna,Johnson:2018hrx} and the USDB $1d_{5/2}-2s_{1/2}-1d_{3/2}$ \cite{Brown:2006gx} and GXPF1 $1f_{7/2}-2p_{3/2}-2p_{1/2}-1f_{5/2}$ \cite{Honma:2002pr} interactions, respectively. Harmonic oscillator bases with oscillator parameters $b$ of 1.84 and 1.99 fm for Al and Ti, respectively, were employed in the calculations. 

The effective one-body couplings which reproduce the shell model results are given in Table \ref{tab:effective_op}. We find that the Fermi gas average estimate is $\approx 2-3$ times larger than the shell model estimate, in good agreement with the anapole study. Although we did not carry out the shell model calculation for the heavy nuclei $^{197}$Au and $^{208}$Pb, given that the anapole study observed the overestimation in $^{205}$Tl, it is likely that a similar result would be found for $^{197}$Au and $^{208}$Pb in our case.

The shell model results may still represent an overestimation of the two-nucleon contribution. The shell model wave functions that we employ are constructed in very soft Hilbert spaces which lack the high-momentum modes necessary to properly resolve the strong repulsion of two nucleons at short distance. We find that the two-nucleon operator is sensitive to the short-range nucleon-nucleon physics. Introducing an ad hoc short-range correlation function \cite{Miller:1975hu} in the shell model calculation further reduces the estimated strength of the two-nucleon operator by $\approx50\%$. For this reason, the value of $f^{SI}_\mathrm{eff}$ obtained from the shell model calculation can likely be considered as an upper limit on the strength of the two-nucleon contribution. A complete treatment involving the introduction of effective operators and wave function renormalization to account for the truncated shell model space is beyond the scope of this paper. 

\section{Transition rate including NLO corrections}
\label{sec:transition}
\renewcommand{\arraystretch}{1.5}
\begin{table}[t]
\caption{Capture rate for $^{27}_{13}$Al, $^{48}_{22}$Ti, $^{197}_{79}$Au  and $^{208}_{82}$Pb. The unit is taken to be neV. }
\begin{tabular}{c c c c c}
\hline
 &~$^{27}_{13}$Al~&~$^{48}_{22}$Ti~&~$^{197}_{79}$Au~&~$^{208}_{82}$Pb~ \\
\hline \hline
$\Gamma_{\rm capt}$~[neV] &~$0.463$~&~$1.705$~&~$8.603$~&~$8.853$~\\
\hline
\end{tabular}
\label{capture_rate}
\end{table}
The rate of the coherent $\mu\to e$ conversion process depends on the behavior of the bound muon and outgoing electron. The lepton wave functions are obtained by solving the Dirac equation \cite{Rose1961,Strange1998,Czarnecki:1998iz,Kitano:2002mt},
\begin{align}
W\psi=\bigg[-i\gamma_5\sigma_r\left(\partial_r+\frac{1}{r}-\frac{\beta}{r}K \right)+V(r)+m\beta \bigg]\psi,
\end{align}
with
\begin{align}
\gamma_5&=
\begin{pmatrix}
0 & 1\\
1 & 0
\end{pmatrix},\hspace{1.5cm}
\beta=
\begin{pmatrix}
1 & 0\\
0 & -1
\end{pmatrix},\\
\sigma_r&=
\begin{pmatrix}
{\boldsymbol \sigma}\cdot {\bf \hat{r}} & 0 \\
0 & {\boldsymbol \sigma}\cdot {\bf \hat{r}}
\end{pmatrix},\\
K&=
\begin{pmatrix}
{\boldsymbol \sigma}\cdot {\boldsymbol {\it l}}+1 & 0\\
0 & -({\boldsymbol \sigma}\cdot {\boldsymbol {\it l}}+1)
\end{pmatrix}.
\end{align}
Here, the energy, potential and mass of the leptons are given by $W,~V(r)$ and $m$. $\boldsymbol \sigma$ are the Pauli matrices, $\bf{\hat r}$ is a unit vector in the radial direction, $\boldsymbol{\it l}$ is the orbital angular momentum defined by $\boldsymbol{\it l}=-i {\bf r}\times \nabla$. We define the wave functions as
\begin{align}
\psi=
\begin{pmatrix}
g_{\kappa}(r)\chi^{\mu}_{\kappa}(\theta, \phi)\\
if_{\kappa}(r)\chi^{\mu}_{-\kappa}(\theta, \phi) 
\end{pmatrix}, \label{wavefunction}
\end{align}
where $\mu$ and $\kappa$ represent the eigenvalues of the $z$ component of the total angular momentum $J_z$ and $K$, respectively. The two-component spinors $\chi^{\mu}_{\kappa}$ are the spin-angular functions, with the properties
\begin{align}
\left({\boldsymbol \sigma}\cdot {\boldsymbol {\it l}}+1\right)\chi^{\mu}_{\kappa}&=-\kappa\chi^{\mu}_{\kappa},\\
J_z\chi^{\mu}_{\kappa}&=\mu \chi^{\mu}_{\kappa},\\
\int^{1}_{-1}d\cos\theta\int^{2\pi}_0d\phi~\chi^{\mu\dagger}_{\kappa}\chi^{\mu^{\prime}}_{\kappa^{\prime}}&=\delta^{\mu\mu^{\prime}}\delta_{\kappa\kappa^{\prime}}.
\end{align}
The initial muon state corresponds to the ground state of the muonic atom, implying $\kappa_{\mu}=-1~$. On the other hand, the outgoing electron has two states of $\kappa_e=\pm 1$. Normalization of the bound muon state is defined by
\begin{align}
\int d^3x~\psi^{(\mu)\dagger}_{\kappa,\mu}(x)\psi^{(\mu)}_{\kappa^{\prime}\mu^{\prime}}(x)=\delta_{\mu\mu^{\prime}}\delta_{\kappa\kappa^{\prime}}.
\end{align}
Neglecting nuclear recoil, the final state electron carries energy $W=m_{\mu}-B_{\mu}$, where $B_{\mu}$ is the binding energy of the $1s$ muonic atom. Its wave function is normalized as
\begin{align}
\int d^3x~\psi^{(e)\dagger}_{\kappa,\mu}(x)\psi^{(e)}_{\kappa^{\prime}\mu^{\prime}}(x)=2\pi\delta(W-W^{\prime})\delta_{\mu\mu^{\prime}}\delta_{\kappa\kappa^{\prime}}.
\end{align}

Inserting the expressions of the wave functions into the the spherical polar form of the Dirac equation,  one can obtain
\begin{align}
\frac{d}{dr}
\begin{pmatrix}
g\\
f
\end{pmatrix}=
\begin{pmatrix}
-\frac{\kappa+1}{r} & W-V(r)+m\\
-(W-V(r)-m) & \frac{\kappa-1}{r}
\end{pmatrix}
\begin{pmatrix}
g\\
f
\end{pmatrix}.
\end{align}
Utilizing the shoot-and-match procedure \cite{Silbar:2010wx}, we solve these coupled equations numerically. 

For $\mu \to e$ conversion, the branching ratio is defined by the conversion-to-capture ratio
\begin{align}
B_{\mu\to e}\equiv\frac{\Gamma_{\rm conv}(\mu^- + (A, Z)\to e^-+(A, Z))}{\Gamma_{\rm capt}(\mu^-+(A, Z)\to \nu_{\mu}+(A, Z-1))},
\end{align}
where $A$ and $Z$ are mass and atomic numbers, respectively. The standard muon capture rates 
 $\Gamma_\mathrm{capt} \equiv \kappa_\mathrm{capt} m_\mu^5/v^4$ 
for the nuclei of interest are listed in Table \ref{capture_rate}. Taking into account all the spin configurations of the initial muon and final electron states, 
one can express the 
branching ratio 
as 
\begin{align}
B_{\mu \to e} =  &
\left(\frac{v}{\Lambda}\right)^4 \frac{1}{\kappa_\mathrm{capt}}
\left(\left|\tau^{(+1)} \right|^2+\left|\tau^{(-1)} \right|^2 \right),
\end{align}
in terms of dimensionless overlap integrals
\begin{align}
 \tau^{(-1)}=&\left(C_{DL}+ C_{DR} \right)\tau^{(-1)}_D+ \tau^{(-1)}_{S},\\
\tau^{(+1)}=&\left(C_{DL}- C_{DR} \right)\tau^{(+1)}_D- \tau^{(+1)}_{S},
\end{align}
where the upper index indicates the quantum number  $\kappa_e$. The overlap integrals for the dipole operator are given by
\begin{align}
\tau^{(-1)}_D&=\frac{1}{m_{\mu}^{{3}/{2}}}\int dr~r^2(-E(r))\left(g^{(e)}_{-1}f^{(\mu)}_{-1}+f^{(e)}_{-1}g^{(\mu)}_{-1} \right),\\
\tau^{(+1)}_D&=\frac{i}{m_{\mu}^{{3}/{2}}}\int dr~r^2(-E(r))\left(g^{(e)}_{+1}g^{(\mu)}_{-1}-f^{(e)}_{+1}f^{(\mu)}_{-1} \right).
\end{align}

For the contributions from the scalar operator, we split the overlap integrals into two contributions, momentum-transfer independent ($\tau_{\rho}^{(\pm 1)}$) and dependent ($\tau_f^{(\pm 1)}$) terms
\begin{widetext}
\begin{subequations}
\label{eq:taus} 
\begin{align}
\tau^{(-1)}_{S}&=\frac{1}{2}G_Fm_{\mu}^2\sum_{N=p,n}\Bigg[\left(C^{\rho}_{NL}+ C^{\rho}_{NR}\right)\tau^{(-1)}_{\rho_N}+\left(C_{NL}^f+ C_{NR}^f\right)\tau^{(-1)}_{f_N}\Bigg],\\
\tau^{(+1)}_{S}&=\frac{1}{2}G_Fm_{\mu}^2\sum_{N=p,n}\left[\left(C_{NL}^{\rho}- C_{NR}^{\rho}\right)\tau^{(+1)}_{{\rho}_N}+\left(C_{NL}^{f}- C_{NR}^{f}\right)\tau^{(+1)}_{f_N}\right],
\end{align}
\end{subequations}
\end{widetext}
where 
$C_{N{\alpha}}^{\rho}$
and 
$C_{N{\alpha}}^{f}$
correspond to 
the constant and momentum-dependent parts of the nucleon scalar form factor, respectively:
\begin{widetext}
\begin{subequations}
\begin{eqnarray}
m_{\mu}C^{\rho}_{N \alpha}  &=&  
\left(\sigma_{\pi N} -\frac{3g^2_Am^2_{\pi}}{64\pi f^2_{\pi}}k_Ff^{\rm SI}_{\rm eff} \right)
\left( C_{S\alpha}^{(0)} + \frac{8 \pi}{9} C_{G\alpha} \right) \mp \frac{\delta m_N}{4} 
\left( C_{S\alpha}^{(1)} + \frac{16 \pi}{9} C_{G\alpha} \right) \nonumber\\
&+&   \sigma_s    \, \left( C_{S\alpha}^{(s)}  + \frac{8 \pi}{9} C_{G\alpha} \right)
- \frac{8 \pi}{9} C_{G\alpha} m_N,\\
m_{\mu}C^{f}_{N\alpha}   &=&   
-\frac{3g^2_Am^3_{\pi}}{64\pi f^2_{\pi}}\frac{5}{12}\left( C_{S\alpha}^{(0)} + \frac{8 \pi}{9} C_{G\alpha} \right) - \dot{\sigma}_sm^2_{\pi} \left( C_{S\alpha}^{(s)} + \frac{8 \pi}{9} C_{G\alpha} \right). \label{CNf}
\end{eqnarray}
\label{eq:nucleon_level_coeffs}
\end{subequations} 
\end{widetext}
The nucleon-level couplings $C^{\rho,f}_{N\alpha}$ are dimensionless, and the minus (plus) sign in the second term of $C^{\rho}_{N \alpha}$ is for proton (neutron).
The dimensionless overlap integrals $\tau_{\rho_N}^{(\pm1)}$ and $\tau_{f_N}^{(\pm 1)}$ are defined as
\begin{align}
\tau^{(-1)}_{\rho_N(f_N)}&=\frac{1}{m_{\mu}^{5/2}}\int dr~r^2\left(g^{(e)}_{-1}g^{(\mu)}_{-1}-f^{(e)}_{-1}f^{(\mu)}_{-1} \right) \rho_N(f_N),\\
\tau^{(+1)}_{\rho_N(f_N)}&=\frac{i}{m_{\mu}^{5/2}}\int dr~r^2\left(f^{(e)}_{+1}g^{(\mu)}_{-1}+g^{(e)}_{+1}f^{(\mu)}_{-1} \right) \rho_N(f_N),
\end{align}
where $\rho_N(r)$ is the nucleon density and the function $f_N(r)$ is given by\footnote{We have numerically checked that the fourth derivative term is roughly two-orders of magnitude smaller than the second derivative. Therefore, we neglect higher-order derivatives in $f_N(r)$.}
\begin{align}
f_N(r)=-\frac{1}{m^2_{\pi}}\left(\frac{\partial^2}{\partial r^2}+\frac{2}{r}\frac{\partial}{\partial r} \right)\rho_N(r).
\label{eq:fN}
\end{align}
In our analysis, we employ the two-parameter Fermi model for the proton and neutron densities
\begin{equation}
\rho_{N}(r)=\frac{\rho_0}{1+e^{(r-R_{N})/a}},
\end{equation}
the parameters of which are given in \cite{Garcia-Recio:1991ocp} for Al, Au, Pb and \cite{DeVries:1987atn} for Ti. We reproduce these parameters in Table \ref{tab:effective_op}. The nucleon density profiles for Al, Au, and Pb include a separate determination of the neutron density from measurements of pionic atoms, whereas for Ti we have only the proton density, determined from electron scattering. In this case, we assume $R_n=R_p$.

It should be noted that, assuming $m_e=0$, one can obtain relations between two electron states, $g^{(e)}_{-1}=\alpha f^{(e)}_{+1}$ and $f^{(e)}_{-1}=\beta g^{(e)}_{+1}$, with $\alpha/\beta=-1$. We evaluate overlap integrals taking $\alpha=-1$ and $\beta=1$, leading to $\tau_D^{(+1)}=i\tau_D^{(-1)}$ and $\tau_S^{(+1)}=-i\tau_S^{(-1)}$ in the massless limit.\footnote{The overlap integrals defined in \cite{Kitano:2002mt} are obtained by scaling $\tau_D^{(-1)}$ and $\tau_{\rho_N}^{(-1)}$ as
\begin{align}
D=\frac{4}{\sqrt{2}}m^{5/2}_{\mu}\tau_D^{(-1)}, \hspace{0.5cm} S^{(N)}=\frac{1}{2\sqrt{2}}m^{5/2}_{\mu}\tau_{\rho_N}^{(-1)}.
\end{align}}

\renewcommand{\arraystretch}{1.5}
\begin{table}[t]
\caption{The dimensionless overlap integrals $\tau_D^{(\pm1)}$, $\tau^{(\pm 1)}_{\rho_N}$ and $\tau^{(\pm 1)}_{f_N}$ computed for the target nuclei $^{27}_{13}$Al, $^{48}_{22}$Ti, $^{197}_{79}$Au and $^{208}_{82}$Pb.
}
\begin{tabular}{c c c c c }
\hline
 & $^{27}_{13}$Al & $^{48}_{22}$Ti & $^{197}_{79}$Au & $^{208}_{82}$Pb  \\
\hline \hline
$\tau_D^{(-1)}$ & $0.0126$ & $0.0292$ & $0.059$ & $0.057$ \\
$-i \tau_D^{(+1)}$ & $0.0126$ & $0.0293$ & $0.059$  & $0.057$ \\
\hline\hline
$\tau_{\rho_p}^{(-1)}$ & $0.043$ & $0.099$ & $0.148$  & $0.14$ \\
$\tau_{\rho_n}^{(-1)}$ & $0.045$ & $0.117$ & $0.161$  & $0.148$ \\
$\tau_{f_p}^{(-1)}$ & $0.029$ & $0.067$ & $0.030$  & $0.022$ \\
$\tau_{f_n}^{(-1)}$ & $0.030$ & $0.079$ & $0.007$  & $-0.0061$ \\
\hline
$-i \tau_{\rho_p}^{(+1)}$ & $-0.043$ & $-0.099$ & $-0.147$  & $-0.138$ \\
$-i \tau_{\rho _n}^{(+1)}$ & $-0.045$ & $-0.116$ & $-0.159$  & $-0.147$ \\
$-i\tau_{f_p}^{(+1)}$ & $-0.029$ & $-0.067$ & $-0.029$ & $-0.022$ \\
$-i \tau_{f_n}^{(+1)}$ & $-0.030$ & $-0.079$ & $-0.0067$  & $0.0063$ \\
\hline \hline
\end{tabular}
\label{overlap}
\end{table}

\renewcommand{\arraystretch}{1.5}
\begin{table}[t]
\caption{The LO and NLO contributions (defined in Eqs. (\ref{I_LO}) $-$ (\ref{I_2N})) in units of MeV$^2$. The two-nucleon contribution $I_{N,2N}^{\mathrm{NLO}(\pm)}$ is computed with the effective one-body operator strength obtained via Fermi gas average, $f_{\mathrm{eff},\mathrm{FGA}}^{\mathrm{SI}}$ (see Table \ref{tab:effective_op}.)}
\begin{tabular}{c c c c c}
\hline
 & $^{27}_{13}$Al & $^{48}_{22}$Ti & $^{197}_{79}$Au &  $^{208}_{82}$Pb  \\
\hline \hline
$I^{\rm LO (-)}_p$ & $268.49$ & $618.10$ & $924.93$ & $871.15$ \\
$I^{\rm NLO (-)}_{p, {\rm loop}}$ & $-9.65$ & $-22.15$ & $-9.79$ & $-7.23$ \\
$I^{\rm NLO (-)}_{p, {\rm 2N}}$ & $-25.2$ & $-70.84$ & $-123.66$  & $-116.47$ \\
\hline
$I^{\rm LO (-)}_n$ & $280.16$ & $730.48$ & $1003.3$ & $923.59$ \\
$I^{\rm NLO (-)}_{n, {\rm loop}}$  & $-10.01$ & $-26.18$ & $-2.31$ & $2.0$ \\
$I^{\rm NLO (-)}_{n, {\rm 2N}}$ & $-26.30$ & $-83.72$ & $-134.13$ & $-123.48$ \\
\hline\hline
$-i I^{\rm LO (+)}_p$ & $-267.22$ & $-615.24$ & $-918.62$ & $-864.55$ \\
$-i I^{\rm NLO (+)}_{p, {\rm loop}}$ & $9.61$ & $22.05$ & $9.69$ & $7.14$ \\
$-i I^{\rm NLO (+)}_{p, {\rm 2N}}$ & $25.08$ & $70.51$ & $122.81$ & $115.58$ \\
\hline
$-i I^{\rm LO (+)}_n$ & $-278.83$ & $-727.10$ & $-995.90$ & $-915.92$ \\
$-i I^{\rm NLO (+)}_{n, {\rm loop}}$ & $9.96$ & $26.06$ & $2.22$ & $-2.06$ \\
$-i I^{\rm NLO (+)}_{n, {\rm 2N}}$ & $26.18$ & $83.34$ & $133.14$ &  $122.45$ \\
\hline \hline
\end{tabular}
\label{overlap_scalar}
\end{table}

Table \ref{overlap} shows the results of the dimensionless overlap integrals $\tau^{(\pm 1)}_D,~\tau^{(\pm 1)}_{\rho_N}$ and $\tau^{(\pm 1)}_{f_N}$ for $^{27}_{13}$Al, $^{48}_{22}$Ti, $^{197}_{79}$Au and $^{208}_{82}$Pb. 
We find that the values of the momentum-dependent overlap integrals $\tau_{f_N}$ for Al and Ti are roughly a factor of $2/3$ smaller than the corresponding momentum-independent integrals $\tau_{\rho_N}$. In heavier nuclei, $\tau_{f_N}$ is further suppressed relative to $\tau_{\rho_N}$.

In the case of non-zero CLFV couplings to $u$ and $d$ quarks, we can compare the LO contributions with those from the NLO interactions by defining the following quantities: 
\begin{align}
I_{N}^{{\rm LO}(\pm)}&\equiv \sigma_{\pi N} m_{\mu}~\tau^{(\pm1)}_{\rho_N}, \label{I_LO}\\
I_{N, {\rm loop}}^{\rm NLO (\pm)}&\equiv -\frac{3g^2_Am^3_{\pi}m_{\mu}}{64\pi f^2_{\pi}}\frac{5}{12}~\tau^{(\pm 1)}_{f_N}, \label{I_loop}\\
I_{N, {\rm2N}}^{\rm NLO(\pm)}&\equiv -\frac{3g^2_Am^2_{\pi}k_Fm_{\mu}}{64\pi f^2_{\pi}}f^{\rm SI}_{\rm eff}~\tau^{(\pm 1)}_{\rho_N}.\label{I_2N}
\end{align}
The resulting values for the nuclei of interest are presented in Table \ref{overlap_scalar}, where input parameters such as $\sigma_{\pi N}$ and $f^{\rm SI}_{\rm eff}$ are fixed at their central values. As discussed in \cite{Bartolotta:2017mff}, the NLO nucleon interactions bring negative relative contributions compared to the LO value, thereby reducing the overall $\mu\rightarrow e$ decay rate. The NLO loop contribution to the amplitude is less than $5\%$  of the LO contribution, while the two-nucleon interactions roughly amount to $10\%$.  If we assume $C_{S\alpha}^{(1)}/C_{S\alpha}^{(0)} = O(1)$, then isospin-breaking terms give only $1\%$ corrections to $I_N^{\rm LO}$ since $(\delta m/4)/\sigma_{\pi N}\approx 0.01$. 
This assumption, however, may be violated in some underlying new physics model. 

In the case of the strange quark, the NLO correction arises only from the momentum-dependent term, $\tau_{f_N}$. Taking the central value of $\dot{\sigma}_s$, we see that the NLO term could reduce the strange-quark contribution to the CLFV amplitude by $10\%$ in Al and Ti, while the corrections are less than a few $\%$ in Au and Pb. Assuming equivalent Wilson coefficients $C^{(0)}_{S\alpha}=C^{(s)}_{S\alpha}$, the total (LO + NLO) contribution to the CLFV amplitude from the strange quark  is reduced by $\approx 30\%$ in Al and Ti and $\approx 20\%$ in Au and Pb, relative to the contribution from $u,d$ quarks. 

In contrast to the light quarks, the NLO contributions to the gluonic coupling $C_{G\alpha}$ are less than $1\%$ relative to the leading term which is dominated by the nucleon mass. Recalling the relation between the heavy quark $(q=c,b,t)$ Wilson coefficients and the gluonic coupling, Eq. (\ref{heavy_qaurks}), we find that the prefactor of the heavy quark Wilson coefficients in Eq. (\ref{eq:nucleon_level_coeffs}a) is $\approx 20\%$ larger than the prefactor of the isoscalar coupling  $C^{(0)}_{S\alpha}$. Therefore in cases where the CLFV quark couplings are independent of flavor, such as the Higgs-mediated model considered in Sec. \ref{sec:results}, the heavy quarks give the largest contribution to the CLFV amplitude (ignoring the intrinsic gluon coupling and the dipole contribution), though all quark flavors contribute at a similar order.
\begin{figure}[t!]
\begin{center}
\includegraphics[scale=0.4]{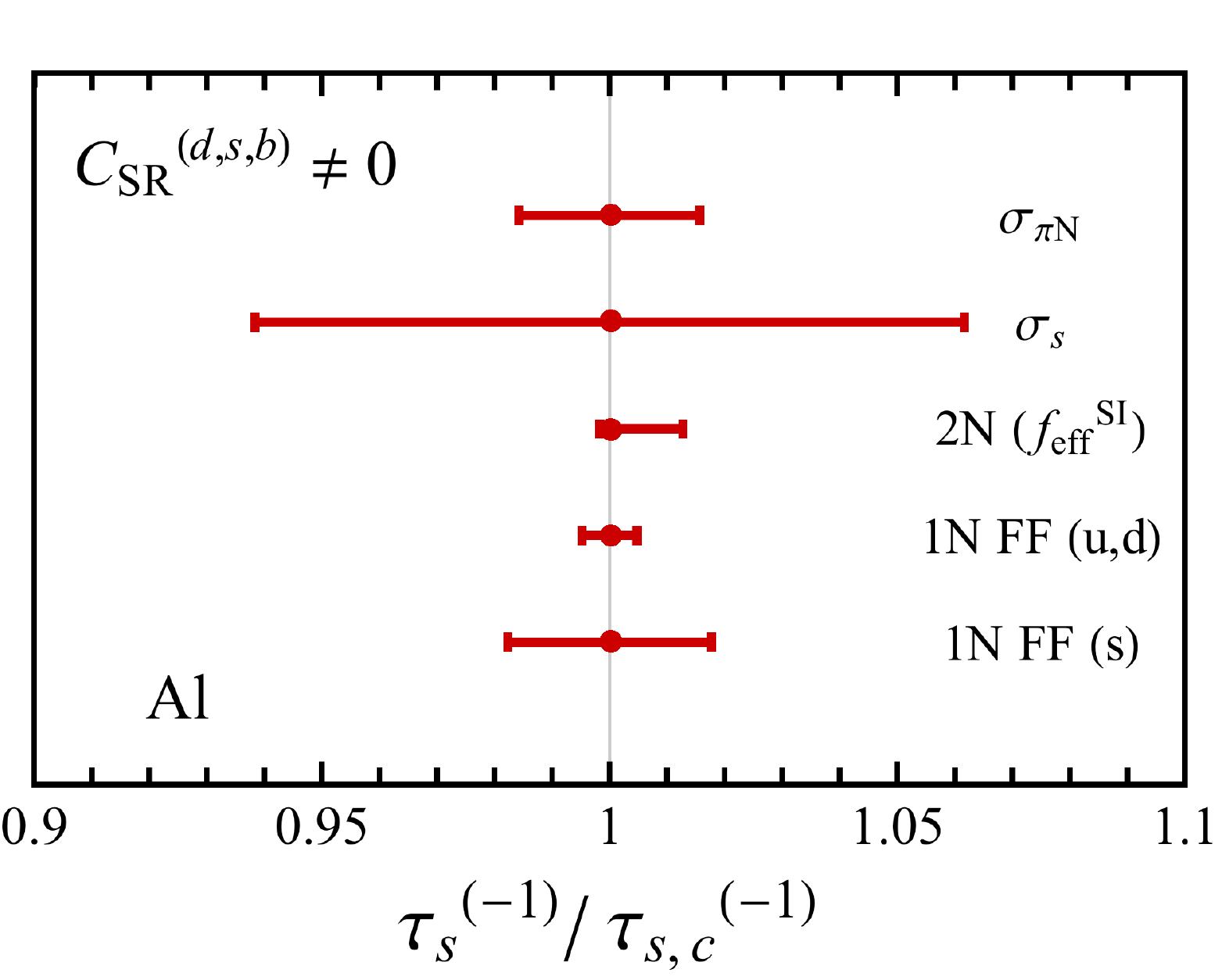}
\includegraphics[scale=0.4]{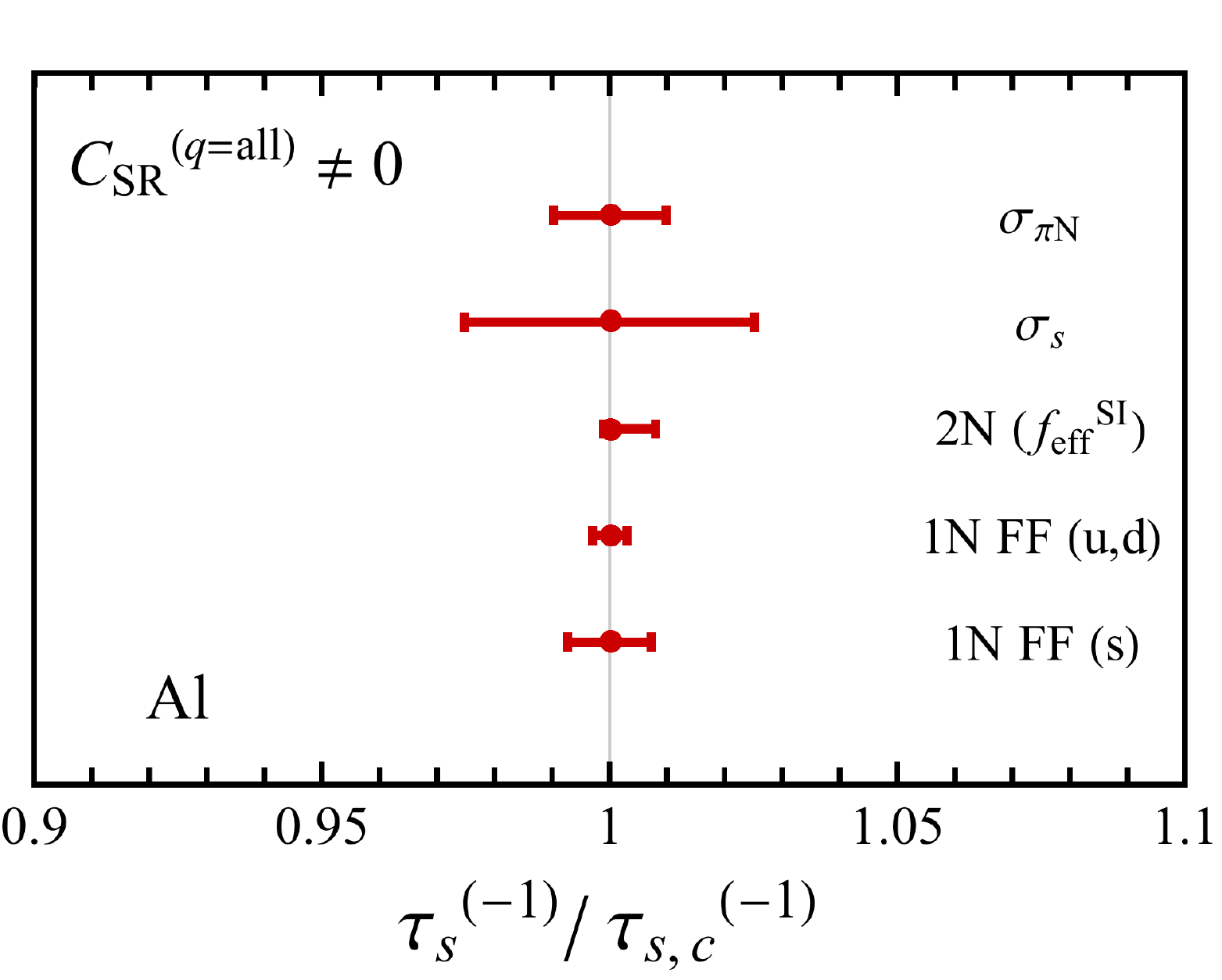}
\end{center}
\caption{Uncertainty budget for the overlap integrals. 
We show the dependence of $\tau^{(-1)}_S/\tau^{(-1)}_{S,c}$ in Al on each parameter: $\sigma_{\pi N}$, $\sigma_s$, $f_{\rm eff}^{\rm SI}$ (denoted by 2N($f_{\rm eff}^{\rm SI}$)) and one-body form factors for $(u,d)$ and $s$ (represented by 1N FF ($u,d$) and ($s$)). Only right-handed down-type operators are nonzero in the upper plot, while the lower plot takes all the right-handed operators to be non-vanishing. In both cases, the nonzero Wilson coefficients are assumed to be equal; that is, $C_{SR}^{(d)}=C_{SR}^{(s)}=C_{SR}^{(b)}$ in the upper plot, and similarly in the lower plot. $\tau^{(-1)}_{S,c}$ is obtained by taking central values of input parameters.}
\label{UCT}
\end{figure}

Fig. \ref{UCT} explores the dependence of the ratio $\tau_S^{(-1)}/\tau_{S,c}^{(-1)}$ on the two input hadronic parameters $\sigma_{\pi N}$ and $ \sigma_s$ as well as the parameters of the NLO nucleon interactions, as discussed in Section~\ref{sec:operator}.\footnote{Since the errors in $\epsilon$, $\delta m$, and $k_F$ are negligible, we do not include them in our analysis.} $\tau_{S,c}^{(-1)}$ is estimated by fixing all the parameters at their central values. Each bar in the plot is obtained by varying one parameter (indicated on the right) in $1~\sigma$ range  while the rest of the parameters are fixed at the central values. The upper panel presents the case with nonzero right-handed down-type operators, while the lower panel includes all right-handed operators.
 
The first two bars in each panel  depict the effect of varying $\sigma_{\pi N}$ and $\sigma_s$. 
The third bar takes into account the error of the effective one-body coupling $f_{\rm eff}^{\rm SI}$ in Table \ref{tab:effective_op}. The lower uncertainty reflects the fact that the central value of $f^{SI}_\mathrm{eff}$ obtained from the Fermi gas average is likely too large by a factor of two. We find a corresponding increase in the value of $\tau_S^{(-1)}$ by $\approx1-2$\% relative to $\tau_{S,c}^{(-1)}$ when $f^{SI}_\mathrm{eff}$ is reduced by half.  
The last two bars correspond to the NLO loop contributions, which are denoted as 1N FF (Form Factor). For the light quark contributions defined by 1N FF ($u,d$), we assign a $50\%$ error to the value of $I^{\rm NLO}_{N, {\rm loop}}$. On the other hand, 1N FF ($s$) represents the variation of $\dot{\sigma}_s$. 

Overall, the variation of the ratio in the lower plot is small compared to the upper plot. This is because, as discussed in \cite{Crivellin:2014cta}, the case has two additional heavy-quark contributions, leading to less impacts from the parameters that we currently focus on.
We find that the scalar contribution is dominantly affected by the uncertainty in $\sigma_s$, which roughly amounts to $\pm 7\%~(3\%)$ for the $C_{SR}^{d,s,b}~(C_{SR}^{q=\rm all})\neq 0$ case. Varying all of these parameters, we see that the deviation from the central value of $\tau_{S,c}^{(-1)}$ is roughly $\pm10\%$ and $\pm5\%$ for the upper and lower case, respectively.

The relative importance of the NLO contributions can depend significantly on the underlying CLFV physics. In particular, when CLFV primarily arises from light quark scalar couplings the NLO contribution can in fact be larger than the LO uncertainty. 
We illustrate this with the following  two  examples:
\begin{itemize}
\item  If only the two lightest quarks contribute, $C_{S\alpha}^{(u)}\approx C_{S\alpha}^{(d)} = O(1)$ and $C_{S\alpha}^{(q=s,c,b,t)}=0$, then the 1-$\sigma$ uncertainty on the LO result for $B_{\mu\to e}({\rm Al})$ is $\pm 13\%$, whereas the NLO contribution reduces the LO branching ratio by roughly 25\%. If the strange quark contributes as well, $C_{S\alpha}^{(s)}= O(1)$, then the LO uncertainty on $B_{\mu\to e}({\rm Al})$ is $\pm 19\%$ while the NLO contribution is $23$\%. In this case, the impact of LO strange quarks is comparable in magnitude to that of the LO light quark uncertainties as well as the NLO light quark contribution.

\item Assuming $C^{(q)}_{S\alpha}\approx 1/y_q$ in both cases, i.e., nonzero $C^{(q=u,d)}_{S\alpha}$ and $C^{(q=u,d,s)}_{S\alpha}$, we see that the negative NLO contributions are $\approx 25\%$, which is consistently larger than the LO uncertainties $\approx \pm 13\%$. This is because the two light-quark contributions dominate the conversion process if the scalar operators are generated with $m_qC^{(q)}_{S\alpha}= O(1)$.
\end{itemize}
This analysis implies two important take-home messages: (1) the overall uncertainty is dominated by the LO amplitude, in particular by the sigma terms; (2) the central value of the NLO corrections could be larger than the uncertainty on the LO term if light quarks ($q=u,d$) have nonzero LFV couplings, making the analysis of NLO effects phenomenologically  relevant. As we find in the next section, the relative significance of the NLO contributions can be diminished in scenarios with large contributions from either gluonic couplings generated by heavy quarks or dipole operators.

It should be noted that in our analysis we have adopted a central value and uncertainty for the two-nucleon contribution based on the Fermi gas model. In light of the nuclear shell model results and their strong dependence on short-range correlations, it is at present impossible to rigorously quantify the uncertainty in the two-nucleon sector. Nonetheless, it is reasonable to regard the values considered here as an upper limit on the relative strength of the two-body contribution (and in turn the NLO contribution).

\section{Phenomenology implications}
\label{sec:results}

We next discuss some phenomenological implications of the improved analysis of transition rates. 

\subsection{Dipole-scalar dominance model}
First, we  consider a restricted EFT setup in which 
only dipole and scalar operators are  generated, assuming that the ratio $C_{D}/C_{S}$ is characterized by a real parameter $r$. 
 As in Ref. \cite{Cirigliano:2009bz}, we assume $C_{DR}=(r/8e)C_{SR}$ with $C_{SR}=C_{SR}^{(d)}=C_{SR}^{(s)}=C_{SR}^{(b)}$, while the rest of the operators are zero.
This scenario may be explicitly realized in some
regions of the R-parity conserving SUSY see-saw parameter
     space~\cite{Kitano:2003wn}
(large $\tan \beta$ and relatively low ``heavy"  Higgs sector)  and  
within R-parity violating SUSY~\cite{Kim:1997rr,Huitu:1997bi,Faessler:1999jf,deGouvea:2000cf}. The nonzero dipole operators generate the $\mu\to e\gamma$ process as well, whose branching ratio is simply expressed as
\begin{align}
B_{\mu\to e\gamma}&\equiv \frac{\Gamma(\mu\to e\gamma)}{\Gamma(\mu\to e\nu_{\mu}\bar{\nu}_e)}\\
&=96\pi^2\left(\frac{v}{\Lambda}\right)^4\left(\left|C_{DR} \right|^2+\left|C_{DL} \right|^2 \right).
\end{align}

\begin{figure}[t!]
\begin{center}
\includegraphics[scale=0.9]{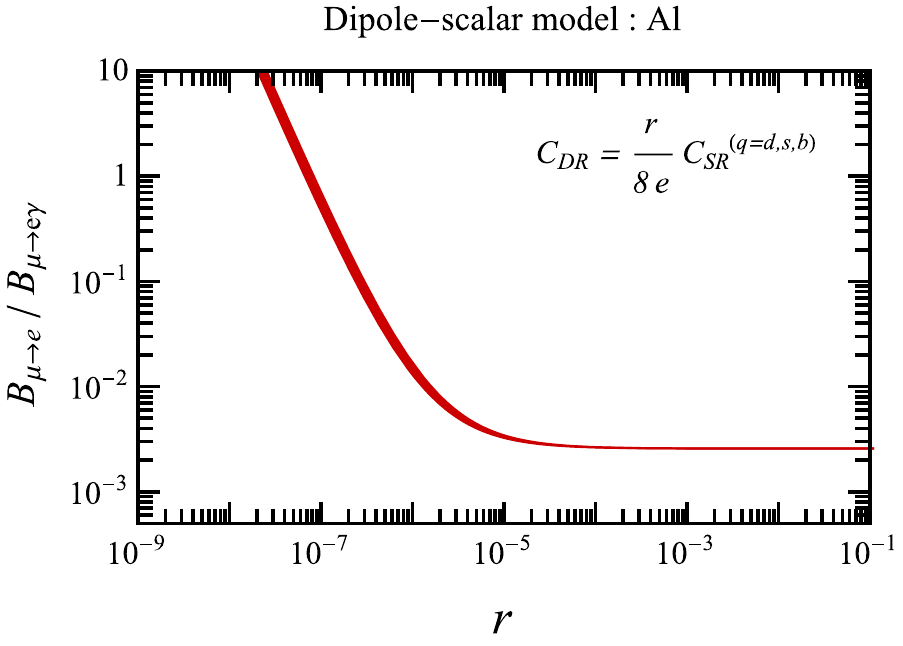}
\includegraphics[scale=0.9]{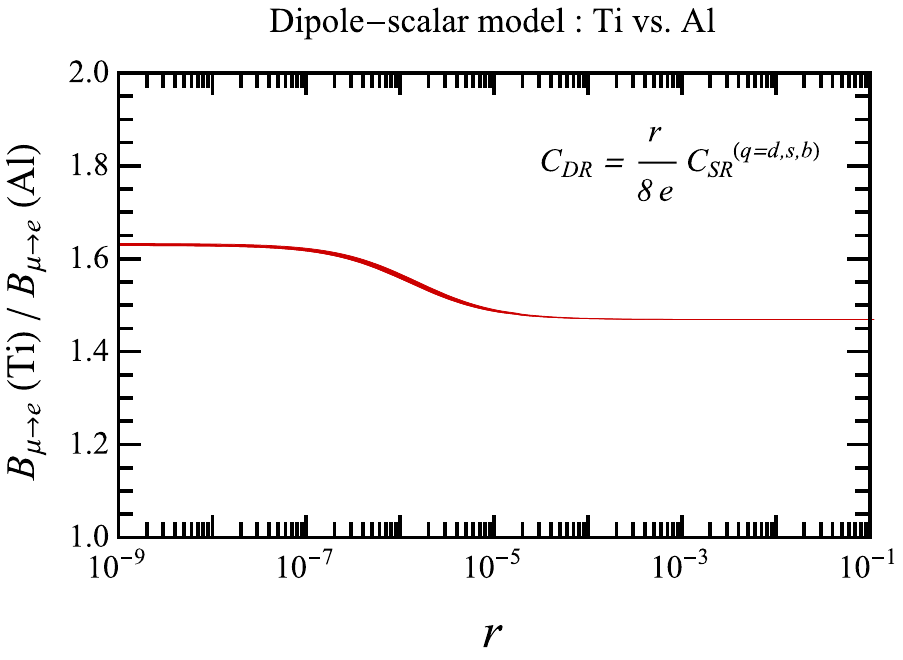}
\end{center}
\caption{Ratios $B_{\mu\to e}({\rm Al})/B_{\mu\to e\gamma}$ (top) and $B_{\mu\to e}({\rm Ti})/B_{\mu\to e}({\rm Al})$~(bottom) against the parameter $r=(8e)C_{DR}/C_{SR}^{(q=d,s,b)}$.}
\label{DS_Al}
\end{figure}

Figure \ref{DS_Al} shows $B_{\mu\to e}({\rm Al})/B_{\mu\to e\gamma}$ in the upper plot and $B_{\mu\to e}({\rm Ti})/B_{\mu\to e}({\rm Al})$ in the lower plot. In this parametrization, the dominant contribution to the $\mu\to e$ conversion switches from the scalar operator to the dipole one around $r \approx 10^{-6}$.  The band in the upper plot is obtained by taking the $1\sigma$ range of the input parameters ($\sigma_{\pi N},\sigma_s, \dot{\sigma}_s, \epsilon, \delta m$), the uncertainty in $f_{\rm eff}^{\rm SI}$ from Table \ref{tab:effective_op}, and a $50\%$ error in the one-body form factor. For example, at $r=10^{-7}$ where the scalar operator dominates $B_{\mu\to e}({\rm Al})$, the uncertainty in $B_{\mu\to e}({\rm Al})/B_{\mu\to e\gamma}$ corresponds to $\pm 20\%$ which can be understood from the analyses of $\tau_S^{(-1)}/\tau_{S,c}^{(-1)}$ in the previous section. On the other hand, taking the ratio of the conversion process between two nuclei, one can see that the uncertainty becomes negligible as seen in the lower plot.  Here, we take the central values of $f_{\rm eff}^{\rm SI}$ for Al and Ti, since we   expect that the uncertainty in the effective one-body coupling is correlated across all isotopes. We see a few $\%$ differences from the results in \cite{Cirigliano:2009bz}.  

However, it should be noted that the ratio $B_{\mu\rightarrow e}$(Ti)/$B_{\mu\rightarrow e}$(Al) is affected by the uncertainty in the overlap integrals due to the neutron densities. As given in Table \ref{tab:effective_op}, the uncertainty in the neutron density parameter $R_n$ determined from pionic atom experiments is roughly 6\% in $^{27}$Al. This propagates to a 5\% error in the neutron overlap integrals for Al. For Ti, in absence of a direct measurement of the neutron density, we employ the same density profile as the measured proton density in Ti. Therefore one would expect a significantly larger uncertainty on the Ti overlap integrals stemming from the neutron density; for example, in the nearby nucleus $^{56}$Fe -- for which a measurement of the neutron density is available -- the difference between using a neutron density measured in pionic atoms compared to assuming identical density profiles for protons and neutrons results in $\approx 7$\% change in the neutron overlap integrals. Accounting for this discrepancy, as well as the uncertainty in the neutron profile parameters,  we assume an 8\% error on the Ti neutron overlap integrals below in Eq. (\ref{eq:ratios}). None of these overlap integral uncertainties are reflected in Figure \ref{DS_Al}, although we expect these errors to be relevant in the scalar-dominated region where $r\lesssim 10^{-5}$.

\subsection{CLFV Yukawa couplings}
Finally, we apply our analysis to the Higgs-mediated CLFV model, where the following Yukawa interactions generate $\mu\to e$ transitions
\begin{align}
{\cal L}=-Y_{e\mu}\bar{e}P_R\mu h -Y_{\mu e}\bar{\mu}P_Re h + {\rm h.c.}.
\end{align}
These Yukawa interactions induce $\mu\to e\gamma$ at one- and two-loop level as discussed in \cite{Harnik:2012pb}.\footnote{For the two-loop contributions, we follow the expressions in \cite{Barr:1990vd, Abe:2013qla}.} On the other hand, the scalar operators arise from a tree-level process mediated by the Higgs particle. The Wilson coefficients are given by
\begin{align}
\frac{1}{\Lambda^2}G_Fm_{\mu}vC_{SR}^{(q)}&=-\frac{1}{m^2_h}Y_{e\mu},\\
\frac{1}{\Lambda^2}G_Fm_{\mu}vC_{SL}^{(q)}&=-\frac{1}{m^2_h}Y_{\mu e},
\end{align}
where the Higgs vacuum expectation value $v=246~$GeV and the Higgs mass $m_h=125~$GeV. 
Note that $C_{S\alpha}^{(q)}$ in this model becomes independent of the label $q$.
Having the parametrization of $C_D=r/(8e)C_S$ that we employ in the previous section, we obtain $r=3.4\times 10^{-6}$ in this model. 

\begin{figure}[t!]
\begin{center}
\includegraphics[scale=0.5]{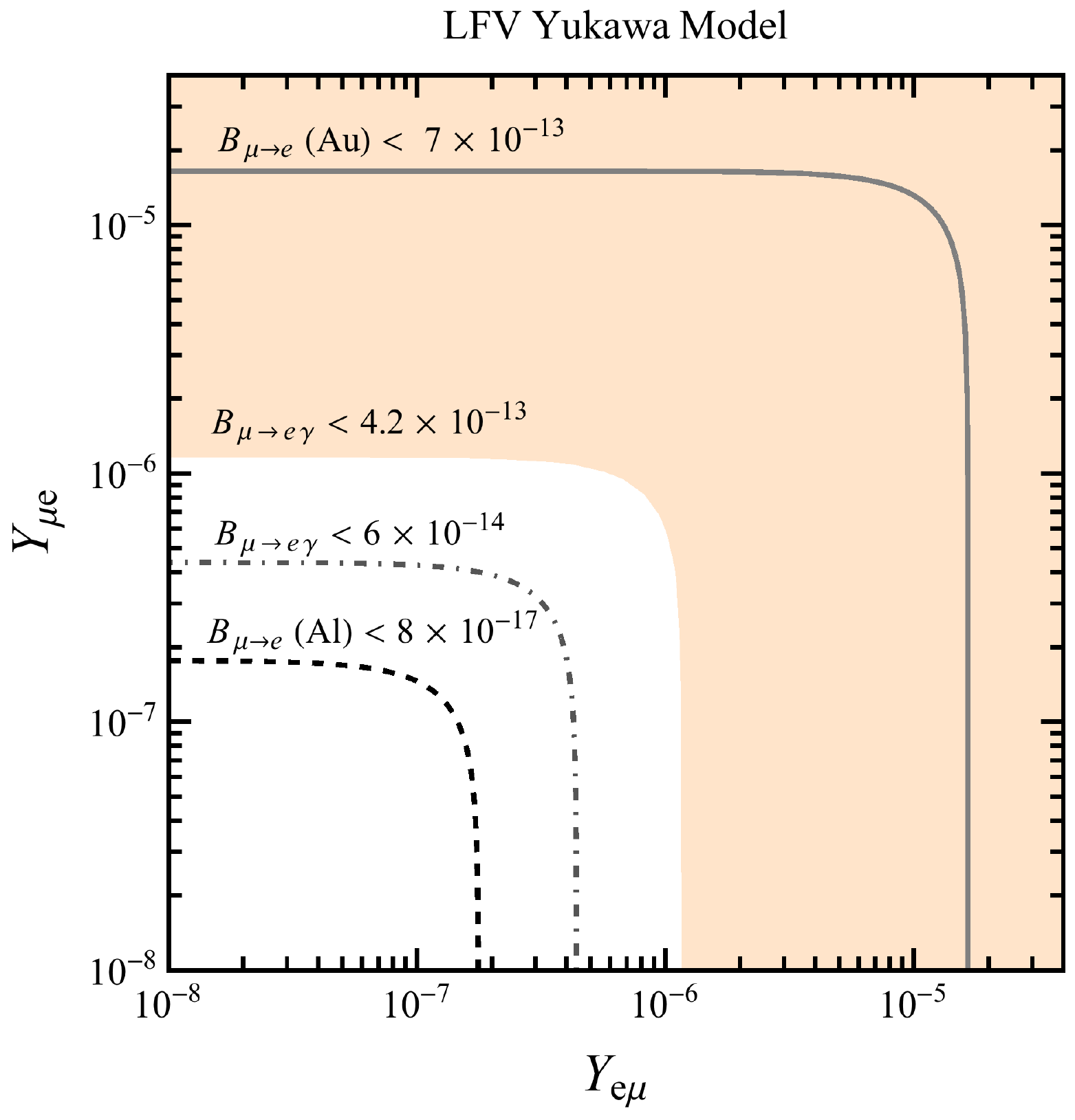}
\end{center}
\caption{Current and prospective limits on $Y_{e\mu}$ and $Y_{\mu e}$ in the CLFV Yukawa model. The gray line is the upper limit from $B_{\mu\to e }$(Au)$<7\times 10^{-13}$, and the orange region is excluded by $B_{\mu\to e \gamma}<4.2\times 10^{-13}$. The expected sensitivities 
at the next-generation experiments are depicted by the black dashed line for $B_{\mu\to e }$(Al)$<8\times 10^{-17}$ and the gray dash-dotted line for $B_{\mu\to e \gamma}<6\times 10^{-14}$, showing the future discovery window.}
\label{LFVYukawa}
\end{figure}

Figure \ref{LFVYukawa} depicts bounds on the CLFV Yukawa couplings $Y_{e\mu}$ and $Y_{\mu e}$. The gray line represents the upper limit on the two couplings from $B_{\mu\to e }$(Au)$<7\times 10^{-13}$. The bound originating from ${\mu\to e\gamma}$ is presented by the orange region, which corresponds to $Y_{e\mu}, Y_{\mu e}\lesssim 10^{-6}$.  
As shown by the black dashed line, 
the next-generation $\mu \to e$ experiments will provide 
a sensitivity to  $Y_{e\mu}$ and  $Y_{\mu e}$ that is stronger than MEG II~\cite{MEGII:2018kmf}
and ten times stronger than current limits.\footnote{To illustrate our point, here 
we take $B_{\mu\to e }$(Al)$<8\times 10^{-17}$ for Mu2e, based on Ref.~\cite{Bernstein:2019fyh}, 
which is slightly weaker than the COMET expected 90\% CL upper limit $2.6\times 10^{-17}$~\cite{COMET:2018auw}.}
The uncertainty resulting from the hadronic input parameters and NLO interactions is not visible  on  the scale of the  plot. 

The plurality of probes, namely $\mu \to e \gamma$ and $\mu \to e$ conversion in possibly more than one target nucleus, 
provides an opportunity to test underlying new physics CLFV mechanisms. 
The minimal Higgs-mediated CLFV scenario considered here produces at low-energy a specific combination 
of scalar and dipole operators,  which leads to  the following pattern of branching ratios:\footnote{The LO results are  $B_{\mu\to e}({\rm Al})/B_{\mu\to e\gamma}=(9.0\pm 0.3)\times 10^{-3}$ and $B_{\mu\to e }({\rm Ti}) /B_{\mu\to e}({\rm Al}) = 1.5\pm0.1$.} 
\begin{subequations}
\begin{eqnarray}
B_{\mu\to e }({\rm Al}) /B_{\mu\to e \gamma} &=& (8.7 \pm 0.3 )\times 10^{-3}
\\
B_{\mu\to e }({\rm Ti}) /B_{\mu\to e}({\rm Al})  &=& 1.5\pm0.1 ~. 
\end{eqnarray}
\label{eq:ratios}
\end{subequations}
Here, we assign a $5 (8) \%$ error to neutron overlap integrals $\tau_{\rho_n}$ and $\tau_{f_n}$ in Al (Ti). While the dominant uncertainty in $B_{\mu\to e }({\rm Al})/B_{\mu\to e \gamma} $ arises from hadronic input parameters, the ratio $B_{\mu\to e }({\rm Ti}) /B_{\mu\to e}({\rm Al})$ is primarily affected by the uncertainties in neutron densities of Al and Ti whereas those of the hadronic parameters are negligible.
Our predictions with quantified uncertainties offer a clean path towards testing 
the Higgs-mediated CLFV scenario, in case of discovery in the next generation experiments.

\section{Conclusions}
\label{sec:conclusion}

In this paper we have studied $\mu \to e$ conversion in nuclei  
within a class of models in which lepton flavor violation is induced dominantly by the  exchange of scalar particles, 
such as the SM Higgs. 
We have estimated  the $\mu \to e$ conversion rate in several nuclei of experimental interest, 
including NLO  effects  in the one-  and two-nucleon interactions in chiral effective theory. 
The one-nucleon effects involve the momentum dependent scalar form factor, and 
we have developed an efficient method to take into account the momentum dependence in the overlap integrals. 
The two-nucleon terms are evaluated both 
by reducing them to one-body terms via an average over a Fermi gas model (as done in Ref.~\cite{Bartolotta:2017mff}) 
and within the nuclear shell model. The two approaches provide  a way to estimate the uncertainty associated with the 
two-body NLO contribution.  
In the process, we correct the result of Ref.~\cite{Bartolotta:2017mff}, finding a smaller effective one-body interaction 
and hence a  smaller impact on the $\mu \to e$ conversion rates. 

For the light-quark scalar operators,
the NLO corrections interfere destructively with the LO terms. 
At the amplitude level, the NLO form-factor contribution is at the 5\% level, while the two-nucleon interactions  amount to about 
10\%,  with a total NLO impact on the decay rates at the 20-30\% level. 
We provide an uncertainty budget for the conversion amplitude at LO and NLO 
and find that the overall uncertainty is dominated by the LO amplitude, in particular by the sigma terms.  Importantly, we find that the central value of the NLO corrections could be larger than the uncertainty on the LO term, making the analysis of NLO effects phenomenologically relevant.

We  studied the  implications of our results for testing scalar-mediated CLFV processes. 
For the phenomenologically interesting  case of  Higgs-mediated CLFV, in which both scalar and dipole operators appear at low-energy, we have shown that:  (i) in the next generation experiments  
$\mu \to e$ conversion will  have stronger sensitivity than $\mu \to e \gamma$ to the CLFV Yukawa couplings $Y_{\mu e, e \mu}$; 
(ii) ratios of branching ratios such as  $B_{\mu\to e }({\rm Al}) /B_{\mu\to e \gamma}$ 
and $B_{\mu\to e }({\rm Al}) /B_{\mu\to e}({\rm Ti})$ can be predicted with quantified uncertainties. 
In particular,  $B_{\mu\to e }({\rm Al}) /B_{\mu\to e \gamma}$ 
is affected by NLO corrections at the same level as LO uncertainties. These ratios offer a clean path towards testing 
the Higgs-mediated CLFV scenario in case of discovery in the next generation experiments.

\begin{acknowledgments}
We are grateful to Anothony Bartolotta, Richard Silbar, Terry Goldman, Anna Hayes-Sterbenz,  Emanuele Mereghetti, 
Martin Hoferichter, Frederic Noel, and Javier Men\'endez 
for valuable discussions. MJRM is supported in part under U.S. Department of Energy contract DE-SC0011095. MJRM was also supported in part under National Science Foundation of China grant No. 19Z103010239. ER is supported in part by the U.S. Department of Energy under grants DE-SC0004658, DE-FOA-0001269 and FWP-NQISCCAWL. VC and KF were supported by the U.S. Department of Energy through the Los
Alamos National Laboratory. KF is also supported by the LANL LDRD Program. Los Alamos
National Laboratory is operated by Triad National Security, LLC, for the National Nuclear Security
Administration of U.S. Department of Energy (Contract No. 89233218CNA000001).
\end{acknowledgments}

\appendix
\section{Transition amplitude in presence of momentum-dependent one body operators}
\label{app:transition}
Let us  consider $\mu \to e$ conversion mediated by a generic  scalar operator of the form:
\begin{align}
O_{L,R} (x) = \bar e (x) P_{L,R} \mu (x) \ \bar q(x) q (x)~. 
\end{align}
In the more commonly studied  case of lepton scattering (electron or neutrino scattering),  one typically uses the mode expansion of the leptonic fields in terms of plane waves. Later on, we will consider this case as a consistency check on the formalism we use. 
However,  it turns out that using plane waves  is not the most convenient choice for the problem of $\mu \to e$ conversion. 
In this case it is more convenient to perform the mode expansion of the leptonic field in terms 
of solutions to the Dirac equation in the Coulomb field of the nucleus.  The muon initial state corresponds to the $1s$ hydrogen-like state, 
while the electron final state corresponds to an outgoing wave in the continuum, with energy given by the muon mass minus the binding energy $B_{\mu}$.

The scalar Hamiltonian at low energy takes the form
\begin{align}
\hat H_{L,R}  \ = \ C_{L,R}  \ \int d^3x \   \bar e (x) P_{L,R} \mu (x) \  \hat{S}  (x) ~, 
\end{align}
where $C_{L,R} \equiv 1/\Lambda_{L,R}^2$ are dimensionful Wilson coefficients and  $\hat{S}   (x)$ is the hadronic realization of the quark scalar density. 
In chiral EFT this operator  contains terms like $\bar N N$, $\pi^+ \pi^-$, etc. 
Its matrix element between free nucleon states takes the form 
\begin{align}
\langle {\bf p}_f | \ \hat S (x)  \ | {\bf p}_i \rangle  \  = \  e^{i x \cdot (p_f - p_i)}   \ g_S ( (p_f - p_i)^2 ) ~, 
\label{eq:SQFT}
\end{align}
where $g_S (q^2)$ is the nucleon scalar form factor (for simplicity we do not display isospin indices here and throughout the discussion). 
In the regime we are working, we could further Taylor expand the scalar form factor as follows:
\begin{align}
g_S (q^2) = g_0 + g_2 q^2 + g_4 (q^2)^2 + ... ~.
\end{align}

Ultimately, we  wish to have a non-relativistic realization of the scalar density operator, to be inserted between nuclear many-body wavefunctions.  
We denote this object by $\hat{ \hat{S}} (x)$ and we determine its form by requiring that the matrix elements within free one-nucleon states are the same in the two representations. Explicitly, 
\begin{align}
\langle {\bf p}_f | \ \hat S (x)  \ | {\bf p}_i \rangle 
  \ = \ 
\int d^3 x_1 \  e^{i {\bf x}_1 \cdot ({\bf p}_i -{\bf p}_f)} \  \hat{\hat{S}} (x; x_1) ~. 
\end{align}
Using Eq.~\eqref{eq:SQFT} and the Taylor expansion of the form factor, one can verify that the desired coordinate-space representation 
of the scalar density involves 
the delta function and its derivatives
\begin{eqnarray}
\hat{\hat{S}} (x; \{x_j\}) &=&  \sum_{i=1}^{A} \ \big[ g_0 \, \delta^{(3)} ({\bf x}_i - {\bf x}) 
\nonumber \\
& -&\ g_2 \, \nabla^2_{x_i}  \delta^{(3)} ({\bf x}_i - {\bf x})  \ + \ ... \big]~, 
\end{eqnarray}
where the summation runs over nucleons. 
%
%

With this result at hand, the conversion amplitude takes the form
\begin{align}
\langle f_g | \ \hat{H}_{L,R}  \ | \mu_{(1s)}  \rangle 
= &     C_{L,R} \ \int d^3x \ \langle e_f |   \,    \bar e (x) P_{L,R} \mu (x) \, | \mu_{(1s)} \rangle \nonumber\\
&\times  
\langle \Psi_0 (\{x_j\}  )| \, \hat{\hat{S}} (x ; \{ x_j\}) \, | \Psi_0 (\{x_j\}) \rangle~, 
\label{eq:amp1}
\end{align}
where $| \Psi_0  \rangle$ is the nuclear many-body ground state wavefunction. 
The leptonic matrix element is expressed in terms of the solutions of the Dirac equation $\psi^{(e),(\mu)} (x)$: 
\begin{align}
 \langle e_f |   \,    \bar e (x) P_{L,R} \mu (x) \, | \mu_{(1s)} \rangle = \bar \psi^{(e)} (x) P_{L,R}  \psi^{(\mu)} (x)   ~.
\label{eq:amp2}
\end{align}
The electron wavefunction  has  an oscillatory behavior with frequency set by  $m_\mu - B_{\mu}$, but it is not a plane wave.   
For the nuclear part of the matrix element one has 
\begin{align}
&\langle \Psi_0 (\{x_j\}  )| \, \hat{\hat{S}} (x ; \{ x_j\}) \, | \Psi_0 (\{x_j\}) \rangle \  \nonumber\\
& = \ 
\int d^3 x_1 ... d^3x_A \ 
\Psi_0^* (x_1, ..., x_A)  \ \hat{\hat{S}} (x; \{x_j\}) \  \Psi_0 (x_1, ..., x_A)~, 
\label{eq:amp3}
\end{align}
and using the definition of one-body density~\footnote{Note that with this definition the proton and neutron densities are normalized as $\int^{\infty}_0dr~4\pi r^2\rho_{p}(r)=Z$ and $\int^{\infty}_0dr~4\pi r^2\rho_{n}(r)=A-Z$.} 
\begin{align}
\rho^{(1)}  (x) = \int d^3 x_2 .... d^3x_A  \ |   \Psi_0 (x, x_2,  ..., x_A)|^2 ~, 
\end{align}
we arrive at 
\begin{align}
&\langle \Psi_0 (\{x_j\}  )| \, \hat{\hat{S}} (x ; \{ x_j\}) \, | \Psi_0 (\{x_j\}) \rangle \  \nonumber\\
&= \ 
g_0 \, \rho^{(1)} (x)  \ - \ g_2 \, \nabla^2  \rho^{(1)} (x)  \ + g_4 \, \nabla^4 \ \rho^{(1)} (x) \ + \cdots~.
\label{eq:amp4}
\end{align}
and hence 
\begin{align}
&\langle e_f | \ \hat{H}_{L,R}  \ | \mu_{(1s)}  \rangle  
 =  C_{L,R}  \int d^3x \ 
 \bar \psi^{(e)} (x)  P_{L,R} \psi^{(\mu)} (x) 
\nonumber \\
& \times  \left[ g_0 \, \rho^{(1)} (x)   
-  g_2 \, \nabla^2  \rho^{(1)} (x) + 
g_4 \, \nabla^4 \rho^{(1)} +  ... \right] ~.
\label{eq:amp5}
\end{align}
This result, together with the relations $\nabla^2 \rho_N (r) =  (1/r) \partial^2/\partial r^2 (r \rho_N (r))$ and $\nabla^4 \rho_N (r) =  (1/r) (\partial^2/\partial r^2)^2 (r \rho_N (r)) $, 
implies the form of the transition amplitude  used in Eqs.~(\ref{eq:taus})-(\ref{eq:fN}).
%
%

Eq.~\eqref{eq:amp5} involves derivatives acting on the nucleon densities in the nuclear ground state and differs from the 
standard results encountered in the analysis of lepton-nucleus scattering. 
In the case of lepton scattering,  the external leptonic  
wavefunctions are plane waves,  and the analogue  of Eq.~\eqref{eq:amp5} 
becomes   (here ${\bf q}= {\bf p}_f - {\bf p}_i$ is the leptonic momentum transfer) 
\begin{align}
& \langle {\bf p}_f | \ \hat{H}_{L,R}  \ | {\bf p}_i  \rangle  
 =  C_{L,R} \bar u(p_e) P_{L,R} u (p_\mu)  \int d^3x \ e^{i {\bf q} \cdot {\bf x} } 
\nonumber \\
& \times 
 \left[ g_0 \, \rho^{(1)} (x)   
-  g_2 \, \nabla^2  \rho^{(1)} (x) + 
g_4 \, \nabla^4 \rho^{(1)} +  ... \right] ~. 
\end{align}
After  integration by parts one recovers the familiar factorized form 
\begin{align}
\label{eq:amp6}
 \langle {\bf p}_f | \ \hat{H}_{L,R}  \ | {\bf p}_i  \rangle  
 & =  C_{L,R} \bar u(p_e) P_{L,R} u (p_\mu) \, 
 \nonumber \\
&  \times   \ g_S( {\bf q}^2 ) \  \int d^3x \ e^{i {\bf q} \cdot {\bf x} } \,    \rho^{(1)} (x)  ~.
\end{align}

The factorized  form of \eqref{eq:amp6} does not directly apply to 
$\mu \to e$ conversion because the leptonic wavefuctions  
$\psi^{(\mu)}(x)$ and $\psi^{(e)} (x)$ are not plane waves. However, one can perform a Fourier decomposition of $\psi^{(\mu)}(x)$ and $\psi^{(e)} (x)$ and apply \eqref{eq:amp6} to each term in the double Fourier expansion in ${\bf p}_e$ and ${\bf p}_\mu$, leading to 
\begin{align}
\label{eq:amp7}
 & \langle e_f | \ \hat{H}_{L,R}  \ |\mu_{(1s)}  \rangle  
  =  C_{L,R}   
 \int d {\bf p}_e  d {\bf p}_\mu \, 
 \bar u(p_e) P_{L,R} u (p_\mu) \, 
 \nonumber \\
&  \times   \ g_S( {\bf p}_e - {\bf p}_\mu)^2  \  \int d^3x \ e^{i ({\bf p}_e - {\bf p}_\mu) \cdot {\bf x} } \,    \rho^{(1)} (x)  ~.
\end{align}
This form is equivalent to \eqref{eq:amp5} but in practice, as long as $g_S ( {\bf q}^2)$ can be approximated by a few terms in its series expansion around ${\bf q}^2=0$, Eq.~\eqref{eq:amp5} is computationally more convenient than \eqref{eq:amp7}.

\begin{widetext}
\section{Two-nucleon Contribution}
\label{app:two_nucleon}
As given in Eq. \ref{eq:currents} (b), the effective two-nucleon Lagrangian obtained from heavy baryon ChPT is
\begin{equation}
J^{(2)}_{ud,\alpha}(\mathbf{q}_1,\mathbf{q}_2)=-\frac{g_A^2m_{\pi}^2}{4f_{\pi}^2}\rho(\mathbf{k}_1,\mathbf{k}_1',\mathbf{k}_2,\mathbf{k}_2')\;\boldsymbol\tau_1\cdot\boldsymbol\tau_2\;C^{(0)}_{S\alpha}
\label{eq:two_nucleon_lagrangian}
\end{equation}
where we have defined the two-nucleon density
\begin{equation}
\rho(\mathbf{k}_1,\mathbf{k}_1',\mathbf{k}_2,\mathbf{k}_2')\equiv \frac{\boldsymbol{\sigma}_1\cdot\mathbf{q}_1\boldsymbol{\sigma}_2\cdot\mathbf{q}_2}{\left(\mathbf{q}_1^2+m_{\pi}^2\right)\left(\mathbf{q}_2^2+m_{\pi}^2\right)}.
\end{equation}
Transforming to position space and imposing momentum conservation, $\mathbf{q}_1+\mathbf{q}_2+\mathbf{q}=0$, we obtain
\begin{equation}
\begin{split}
\rho(\mathbf{x}_1,\mathbf{x}_1',\mathbf{x}_2,\mathbf{x}_2',\mathbf{q})&=\int\frac{d^3k_1}{(2\pi)^3}\frac{d^3k_2}{(2\pi)^3}\frac{d^3k_1'}{(2\pi)^3}\frac{d^3k_2'}{(2\pi)^3}\rho(\mathbf{k}_1,\mathbf{k}_1',\mathbf{k}_2,\mathbf{k}_2')\;(2\pi)^3\delta(\mathbf{k}_1+\mathbf{k}_2-\mathbf{k}_1'-\mathbf{k}_2'-\mathbf{q})e^{i(\mathbf{k}_1'\cdot\mathbf{x}_1'+\mathbf{k}_2'\cdot\mathbf{x}_2'-\mathbf{k}_1\cdot\mathbf{x}_1-\mathbf{k}_2\cdot\mathbf{x}_2)}\\
&=\frac{1}{8\pi}\delta(\mathbf{x}_1-\mathbf{x}_1')\delta(\mathbf{x}_2-\mathbf{x}'_2)e^{-i\mathbf{q}\cdot\mathbf{R}}\int_0^1d\alpha\;e^{i(\alpha-1/2)\mathbf{q}\cdot\mathbf{r}-\Pi(q,\alpha)r}\Bigg\{-i\alpha\hat{\mathbf{r}}\cdot\boldsymbol{\sigma}_1\mathbf{q}\cdot\boldsymbol{\sigma}_2\\
&+i(1-\alpha)\mathbf{q}\cdot\boldsymbol{\sigma}_1\hat{\mathbf{r}}\cdot\boldsymbol{\sigma}_2-\frac{1}{3}\boldsymbol{\sigma}_1\cdot\boldsymbol{\sigma}_2\frac{1}{r}\left[2-r\Pi(q,\alpha)\right]\\
&+\alpha(1-\alpha)\frac{1}{\Pi(q,\alpha)}\mathbf{q}\cdot\boldsymbol{\sigma}_1\mathbf{q}\cdot\boldsymbol{\sigma}_2+\sqrt{\frac{8\pi}{15}}\frac{1}{r}\left[\boldsymbol{\sigma}_1\otimes\boldsymbol{\sigma}_2\right]_2\odot Y_2(\hat{\mathbf{r}})\left[1+r\Pi(q,\alpha)\right]\Bigg\},
\label{eq:two_body_master}
\end{split}
\end{equation}
where we have defined
\begin{equation}
\Pi^2(q,\alpha)\equiv \alpha(1-\alpha)q^2+m_{\pi}^2,
\end{equation}
and introduced the relative and center-of-mass coordinates
\begin{equation}
\begin{split}
\mathbf{r}&=\mathbf{x}_1 - \mathbf{x}_2,\;\mathbf{R}\equiv \frac{1}{2}\left(\mathbf{x}_1+\mathbf{x}_2\right).
\end{split}
\end{equation}
In the $\mathbf{q}\rightarrow 0$ limit, the Feynman parameter integral becomes trivial, $\Pi(q,\alpha)\rightarrow m_{\pi}$, and the two-body charge becomes
\begin{equation}
\rho(\mathbf{x}_1,\mathbf{x}_1',\mathbf{x}_2,\mathbf{x}_2',0)=\frac{1}{8\pi}\delta(\mathbf{x}_1-\mathbf{x}_1')\delta(\mathbf{x}_2-\mathbf{x}_2')\frac{1}{r}\left\{\frac{1}{3}F_1(r/m_{\pi})\boldsymbol{\sigma}_1\cdot\boldsymbol{\sigma}_2+\sqrt{\frac{8\pi}{15}}F_2(r/m_{\pi})Y_2(\hat{\mathbf{r}})\odot\left[\boldsymbol{\sigma}_1\otimes\boldsymbol{\sigma}_2\right]_2\right\},
\end{equation}
where the form factors are given by
\begin{equation}
F_1(x)\equiv e^{-x}(x-2),\;F_2(x)\equiv e^{-x}(x+1).
\end{equation}
This is a familiar result from studies of $0\nu\beta\beta$-decay (e.g. \cite{Prezeau:2003xn}), where the leading long-range contribution is due to two-pion exchange. In that case, the three-momentum transfer is small compared to the pion mass, and it is justified to work in the limit $\mathbf{q}\rightarrow 0$. In coherent $\mu\rightarrow e$ conversion $q\approx m_{\mu}\approx m_{\pi}$, and we must work at finite $\mathbf{q}$. We find that the strength of the two-nucleon operator at $q\approx m_{\mu}$ is reduced by roughly 40\% relative to the $q=0$ value.

The final step in our evaluation is to Fourier transform with respect to $\mathbf{q}$ and then to multipole-project the resulting two-nucleon operator. We are interested in the coherent contribution, the $J=0$ multipole
\begin{equation}
\begin{split}
\mathcal{M}^{(2)}_{J=0,M=0}(q_T)&=\int d^3x \;j_0(q_T x)Y_{0,0}(\hat{\mathbf{x}})\rho(\mathbf{x}_1,\mathbf{x}_1',\mathbf{x}_2,\mathbf{x}_2',\mathbf{x})\\
&=\frac{1}{4\pi}\delta(q_T-q)\int d\Omega_q\; Y_{0,0}(\hat{\mathbf{q}})\rho(\mathbf{x}_1,\mathbf{x}_1',\mathbf{x}_2,\mathbf{x}_2',\mathbf{q}).
\end{split}
\end{equation}
The resulting operator 
\begin{equation}
\mathcal{O}^{(2)}(q_T)=-\frac{g_A^2m_{\pi}^2}{4f_{\pi}^2}\mathcal{M}^{(2)}_{J=0,M=0}(q_T)\;\boldsymbol{\tau}_1\cdot\boldsymbol{\tau}_2\;C_{S\alpha}^{(0)}
\end{equation}
can be evaluated, for example, using the $J=0$, $T=0$ two-body density matrix obtained from the shell model wave function. In order to compare to the one-body average, we equate
\begin{equation}
\braket{J_i||\mathcal{O}^{(2)}(q_T)||J_i}=-\frac{3g_A^2m_{\pi}^2k_F}{64\pi f_{\pi}^2}f_\mathrm{eff}^{SI}\braket{J_i||M_0(q_T)||J_i},
\label{eq:2-body_1-body}
\end{equation}
where $M_{0,0}(q)=\sum_{i=1}^A j_0(q r_i)Y_{0,0}(\hat{\mathbf{r}}_i)$ is the standard one-body charge multipole operator, and $\braket{J_i||\mathcal{O}||J_i}$ denotes a reduced matrix element. Note that for the purpose of extracting $f^{SI}_\mathrm{eff}$ from the shell model calculation, the one-body operator on the right-hand side of Eq. \ref{eq:2-body_1-body} is evaluated with the nuclear shell model density. In general, the scalar single-nucleon density obtained from the nuclear shell model differs from the 2-parameter Fermi function obtained in electron scattering experiments. For example in $^{27}$Al,
\begin{equation}
\braket{J_i||M_0(q_T)||J_i}_\mathrm{2pF}=11.63\pm 0.23,
\end{equation}
where the uncertainty is due to the uncertainty in the neutron density, and
\begin{equation}
\braket{J_i||M_0(q_T)||J_i}_\mathrm{NSM}=11.81.
\end{equation}

\section{Effective One-body Operators}
\label{app:1body_eff}
As discussed in Sec. \ref{sec:operator} and App. \ref{app:two_nucleon}, there is a two-nucleon diagram that contributes to coherent $\mu\rightarrow e$ conversion at NLO. This operator may be replaced by an effective one-body operator by averaging the two-nucleon operator over a degenerate Fermi gas model of the target nucleus. In general, a one-body effective operator can be obtained from a two-body operator by performing a mean-field-like sum over direct and exchange terms
\begin{equation}
\braket{\alpha|\mathcal{O}^{(1)}|\beta}\equiv\sum_{\gamma}\braket{\alpha\gamma|\mathcal{O}^{(2)}|\beta\gamma}-\braket{\alpha\gamma|\mathcal{O}^{(2)}|\gamma\beta},
\end{equation}
where $\ket{\alpha}$ is a single-particle state and the sum runs over all occupied states. In the non-relativistic Fermi gas model of the nucleus, the single-particle states are direct products of momentum, spin, and isospin states
\begin{equation}
\ket{\alpha}=\ket{\mathbf{p}(\alpha)}\otimes\ket{\frac{1}{2}m_s(\alpha)}\otimes\ket{\frac{1}{2}m_t(\alpha)},
\end{equation}
allowing the sums over the three components to be performed independently.

The resulting spin-independent and spin-dependent one-body effective operators depend on the magnitude of the three-momentum transfer $q_T=|\mathbf{k}_f-\mathbf{k}_i|$, the average nucleon momentum $k=\frac{1}{2}|\mathbf{k}_i+\mathbf{k}_f|$, the pion mass $m_{\pi}$ and the nuclear Fermi momentum $k_F$. Introducing the dimensionless variables
\begin{equation}
\bar{q}=\frac{q_T}{k_F},\;\bar{k}=\frac{k}{k_F},\;\bar{m}=\frac{m_{\pi}}{k_F},
\end{equation}
the resulting spin-independent and spin-dependent form factors are 
\begin{equation}
\begin{split}
f^{SI}(\bar{q},\bar{k})&=\frac{2}{\pi}\int_{-1/2}^{1/2}d\beta\left[2\left(1+\frac{-\beta\bar{k}\cdot\bar{q}+\beta^2\bar{q}^2}{\bar{k}^2-2\beta\bar{k}\cdot\bar{q}+\beta^2\bar{q}^2}\right)\right.\\
&-\left(\frac{4(\frac{1}{4}-\beta^2)\bar{q}^2+3\bar{m}^2}{\sqrt{(\frac{1}{4}-\beta^2)\bar{q}^2+\bar{m}^2}}\right)\left(\arctan\left(\frac{1+\sqrt{\bar{k}^2-2\beta\bar{k}\cdot\bar{q}+\beta^2\bar{q}^2}}{\sqrt{(\frac{1}{4}-\beta^2)\bar{q}^2+\bar{m}^2}}\right)+\arctan\left(\frac{1-\sqrt{\bar{k}^2-2\beta\bar{k}\cdot\bar{q}+\beta^2\bar{q}^2}}{\sqrt{(\frac{1}{4}-\beta^2)\bar{q}^2+\bar{m}^2}}\right)\right)\\
&+\frac{1}{2\sqrt{\bar{k}^2-2\beta\bar{k}\cdot\bar{q}+\beta^2\bar{q}^2}}\left(1+2\bar{m}^2+\left(\frac{3}{4}-4\beta^2\right)\bar{q}^2-\bar{k}^2+2\beta\bar{k}\cdot\bar{q}\right.\\
&\left.\left.+\beta\frac{\left(1+\frac{1}{4}\bar{q}^2+\bar{m}^2+\bar{k}^2-2\beta\bar{k}\cdot\bar{q}\right)\left(\bar{k}\cdot\bar{q}-\beta\bar{q}^2\right)}{\bar{k}^2-2\beta\bar{k}\cdot\bar{q}+\beta^2\bar{q}^2}\right)\log\left(\frac{1+2\sqrt{\bar{k}^2-2\beta\bar{k}\cdot\bar{q}+\beta^2\bar{q}^2}+\bar{k}^2-2\beta\bar{k}\cdot\bar{q}+\frac{1}{4}\bar{q}^2+\bar{m}^2}{1-2\sqrt{\bar{k}^2-2\beta\bar{k}\cdot\bar{q}+\beta^2\bar{q}^2}+\bar{k}^2-2\beta\bar{k}\cdot\bar{q}+\frac{1}{4}\bar{q}^2+\bar{m}^2}\right)\right],
\end{split}
\end{equation}
\begin{equation}
\begin{split}
f^{SD}(\bar{q},\bar{k})&=-\frac{2}{\pi}\int_{-1/2}^{1/2}d\beta\frac{1}{\sqrt{\bar{k}^2-2\beta\bar{k}\cdot\bar{q}+\beta^2\bar{q}^2}}\left[\frac{1}{\sqrt{\bar{k}^2-2\beta\bar{k}\cdot\bar{q}+\beta^2\bar{q}^2}}\right.\\
&\left.-\frac{1+\bar{m}^2+\bar{k}^2-2\beta\bar{k}\cdot\bar{q}+\frac{1}{4}\bar{q}^2}{4\left(\bar{k}^2-2\beta\bar{k}\cdot\bar{q}+\beta^2\bar{q}^2\right)}\log\left(\frac{1+2\sqrt{\bar{k}^2-2\beta\bar{k}\cdot\bar{q}+\beta^2\bar{q}^2}+\bar{k}^2-2\beta\bar{k}\cdot\bar{q}+\frac{1}{4}\bar{q}^2+\bar{m}^2}{1-2\sqrt{\bar{k}^2-2\beta\bar{k}\cdot\bar{q}+\beta^2\bar{q}^2}+\bar{k}^2-2\beta\bar{k}\cdot\bar{q}+\frac{1}{4}\bar{q}^2+\bar{m}^2}\right)\right].
\end{split}
\end{equation}
These functions depend not only on the magnitude of the dimensionless momentum transfer $\bar{q}$ and average momentum $\bar{k}$, but on their relative angle. Fortunately, for the physically relevant values of these momenta, $f^{SI}$ and $f^{SD}$ do not vary significantly over the range of possible angular values. Therefore we may replace each function by its angular average. The angle-averaged functions then depend only on the magnitude of the momentum transfer $\bar{q}$ and the average momentum $\bar{k}$. For a given nucleus, the three-momentum transfer in coherent $\mu\rightarrow e$ conversion satisfies
\begin{equation}
q_T^2 \simeq \frac{M_T}{m_{\mu}+M_T}\left[\left(m_{\mu}-E_{\mu}^\mathrm{bind}\right)^2-m_e^2\right],
\label{eq:q_val}
\end{equation}
where $M_T$ is the mass of the target nucleus and $E_{\mu}^\mathrm{bind}$ is the (positive) binding energy of the captured muon. Fixing the value of $q_T$ for a given nucleus, $f^{SI}$ and $f^{SD}$ become functions of the dimensionless average nucleon momentum $\bar{k}$, as shown in Fig. \ref{fig:1body_f_functions}. In order to recover a local one-body effective operator, we replace these slowly-varying functions of $\bar{k}$ by a constant, weighting our average by the single-nucleon momentum probability distribution. The resulting momentum-averaged values $f^{SI}_\mathrm{eff}$ and $f^{SD}_\mathrm{eff}$ are shown in Fig. \ref{fig:1body_f_functions}. Table \ref{tab:effective_op} reports the values of $f^{SI}_\mathrm{eff}$ for the four nuclei of interest as well as the physical parameter values employed in each calculation.
\begin{figure}
\centering
\subfloat[]{
\includegraphics[scale=0.66]{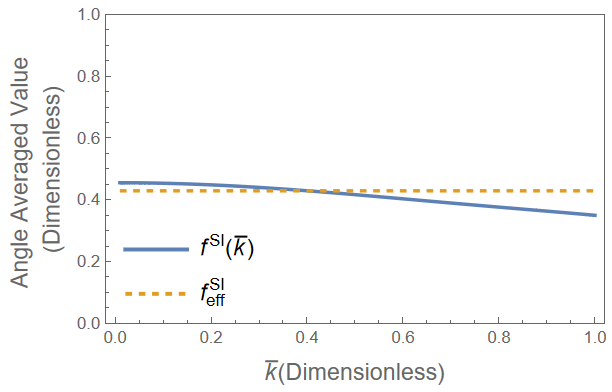}
}
\subfloat[]{
\includegraphics[scale=0.66]{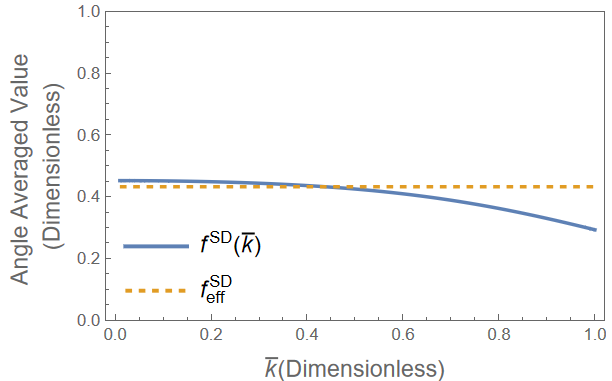}
}
\caption{Left (right): Angle-averaged value of $f^{SI}$ ($f^{SD}$) and its constant approximation $f_\mathrm{eff}^{SI}$ ($f_\mathrm{eff}^{SD}$) as a function of the dimensionless average momentum $\bar{k}$ for the case of $^{27}$Al.}
\label{fig:1body_f_functions}
\end{figure}
\end{widetext}

\bibliography{references}

\end{document}